\documentclass[fleqn,usenatbib]{mnras}

\usepackage{newtxtext,newtxmath}

\usepackage[T1]{fontenc}
\usepackage{appendix}
\usepackage{siunitx}
\usepackage{tablefootnote}

\DeclareRobustCommand{\VAN}[3]{#2}
\let\VANthebibliography\thebibliography
\def\thebibliography{\DeclareRobustCommand{\VAN}[3]{##3}\VANthebibliography}

\usepackage{graphicx}	
\usepackage{amsmath}	
\usepackage{multirow}

\newcommand{\arcsecs}{\mbox{$^{\prime\prime}$}}
\newcommand{\Msolar}{\mbox{$\rm M_{\odot}\,$}}
\def\gs{\mathrel{\raise0.35ex\hbox{$\scriptstyle >$}\kern-0.6em \lower0.40ex\hbox{{$\scriptstyle \sim$}}}}
\def\ls{\mathrel{\raise0.35ex\hbox{$\scriptstyle <$}\kern-0.6em \lower0.40ex\hbox{{$\scriptstyle \sim$}}}}

\title[MIDIS: Strong H$\beta$+O\textsc{iii} Line Emitters]{MIDIS: Strong H$\beta$+[O\textsc{iii}] Line Emitters at $z\gs 9$}
\author[T. R. Greve et al.]
{
Thomas R. Greve,$^{1,2}$\thanks{E-mail: tgreve@space.dtu.dk},
Steven Gillman,$^{1,2}$,
Pierluigi Rinaldi${^3}$,
Iris Jermann$^{1,2}$,
Jens Melinder${^4}$,
\newauthor
G{\"o}ran {\"O}stlin$^{4}$,
Pablo G. P\'erez-Gonz\'alez$^{5}$,
Luis Colina${^5}$,
Fabian Walter$^{6}$,
Javier Álvarez-Márquez$^{5}$,
\newauthor
Martin J.~Ward$^{7}$, 
Alejandro Crespo G\'omez$^{4}$,
John P.~Pye$^{8}$,
Tuomo V.~Tikkanen$^{8}$,
Edoardo Iani$^{9}$,
\newauthor
Roman A. Meyer$^{10}$,
Leindert A. Boogaard$^{11}$,
Jens Hjorth$^{12}$,
Danial Langeroodi$^{12}$,
Paul van der Werf $^{11}$,
\newauthor
Sarah E. I. Bosman$^{6,13}$,
Karina I. Caputi$^{14}$,
Luca Costantin$^{5}$,
Marianna Annunziatella$^{5}$,
Arjan Bik$^{4}$,
\newauthor
\'Alvaro Labiano$^{15}$,
Thomas Henning$^{16}$
\medskip\\
$^{1}$\hypertarget{DAWN}{Cosmic Dawn Center (DAWN)}\\
$^{2}$\hypertarget{DTU}{DTU-Space, Technical University of Denmark, Elektrovej 327, DK-2800 Kgs. Lyngby, Denmark}\\
$^{3}$\hypertarget{}{Space Telescope Science Institute, 3700 San Martin Drive, Baltimore, Maryland 21218, USA}\\
$^{4}$\hypertarget{}{Department of Astronomy, Stockholm University, Oscar Klein Centre, AlbaNova University Centre, 106 91 Stockholm, Sweden}\\
$^{5}$\hypertarget{CAB}{Centro de Astrobiolog\'{\i}a, CSIC-INTA, Ctra. de Ajalvir km 4, Torrej\'on de Ardoz, E-28850, Madrid, Spain}\\
$^{6}$\hypertarget{}{Max Planck Institut f\"ur Astronomie, K\"onigstuhl 17, D-69117, Heidelberg, Germany}\\
$^{7}$\hypertarget{}{Centre for Extragalactic Astronomy, Department of Physics, Durham University, South Road, Durham DH1 3LE, UK}\\
$^{8}$\hypertarget{}{School of Physics \& Astronomy, Space Park Leicester, University of Leicester, 92 Corporation Road, Leicester LE4 5SP, UK}\\
$^{9}$\hypertarget{}{Institute of Science and Technology Austria (ISTA), Am Campus 1, 3400 Klosterneuburg, Austria}\\
$^{10}$\hypertarget{}{Department of Astronomy, University of Geneva, Chemin Pegasi 51, CH-1290 Versoix, Switzerland}\\
$^{11}$\hypertarget{}{Leiden Observatory, Leiden University, PO Box 9513, NL-2300 RA Leiden, The Netherlands}
$^{12}$\hypertarget{}{DARK, Niels Bohr Institute, University of Copenhagen, Jagtvej 155A, 2200 Copenhagen, Denmark}\\
$^{13}$\hypertarget{}{Institute for Theoretical Physics, Heidelberg University, Philosophenweg 12, D–69120 Heidelberg, Germany}\\
$^{14}$\hypertarget{}{Kapteyn Astronomical Institute, University of Groningen, P.O. Box 800, 9700AV Groningen, The Netherlands}\\
$^{15}$\hypertarget{}{Telespazio UK for the European Space Agency, ESAC, Camino Bajo del Castillo s/n, 28692 Villanueva de la Ca\~nada, Spain}\\
$^{16}$\hypertarget{}{Max-Planck-Institut f\"ur Astronomie (MPIA), K\"onigstuhl 17, 69117 Heidelberg, Germany}\\
}
\date{Accepted XXX. Received YYY; in original form ZZZ}

\pubyear{2026}

\begin{document}
\label{firstpage}
\pagerange{\pageref{firstpage}--\pageref{lastpage}}
\maketitle

\begin{abstract}
We present a search for strong H$\beta$+[O\textsc{iii}] line emitters across
the redshift range $z=9.4-11.3$ in the Hubble Ultra Deep Field using ultra-deep
MIRI/F560W imaging ($28.59\,{\rm mag}$, AB, 5-$\sigma$ point-source
sensitivity) from the MIRI Deep Imaging Survey (MIDIS). Three galaxies are
identified via pronounced F560W flux excesses relative to their underlying
continuum, consistent with strong rest-frame optical line emission. From
spectral energy distribution modelling we derive rest-frame
H$\beta$+[O\textsc{iii}] equivalent widths in the range $\sim
600-1300\,\si{\angstrom}$ (median value $\simeq
1260^{+327}_{-259}\,\si{\angstrom}$), placing these objects among the most
extreme nebular line emitters known at these epochs. We combine our MIDIS
sources with a compiled literature sample of 16 spectroscopically confirmed
galaxies at $z \geq 9$ with published H$\beta$+[O\textsc{iii}] equivalent width
measurements and associated physical properties. We find a median ${\rm
EW}^{\rm H\beta+[O\textsc{iii}]}_{\rm rest} \simeq
1318^{+544}_{-385}\,\si{\angstrom}$, similar to values observed in star-forming
galaxies at $z \sim 6-9$. We find no evidence for a steep increase nor a
systematic decline in H$\beta$+[O\textsc{iii}] equivalent widths beyond $z \sim
9$. Binning our combined $z\geq 9$ sample in UV luminosity, we find higher
equivalent widths for the more UV luminous systems, which is qualitatively
consistent with trends reported at $z=6-9$. We do not find a statistically
significant anti-correlation between H$\beta$+[O\textsc{iii}] equivalent width
and stellar mass within our $z\geq 9$ sample. However, a log-linear fit to the
data suggests a trend broadly consistent with the anti-correlation observed at
lower redshift. We place a first constraint on the H$\beta$+[O\textsc{iii}]
line luminosity function at $z\simeq 9-11$ ($\Phi \sim 10^{-3.4}\,{\rm
Mpc^{-3}\,dex^{-1}}$ at $\log(L_{\rm H\beta+[O\textsc{iii}]}/{\rm erg\,s^{-1}})
= 42.5$), which is consistent with a general decline compared to spectroscopic
determinations of the luminosity function at $z\simeq 7-8$.  For our MIDIS
sources, we derive ionising photon production efficiencies in the range
$\log(\xi_{\rm ion}/{\rm Hz\,erg^{-1}}) = 25.1-25.4$. Using our combined $z\geq
9$ sample, we have examined scaling relations between $\xi_{\rm ion}$ and
H$\beta$+[O\textsc{iii}] equivalent width, UV luminosity, and UV continuum
slope. We find statistically significant correlation between $\xi_{\rm ion}$
and ${\rm EW}_{\rm rest}^{\rm H\beta+[O\textsc{iii}]}$ and between $\xi_{\rm
ion}$ and $\beta$, which are also consistent with those observed at $z\simeq
5-9$.  No significant correlation of $\xi_{\rm ion}$ with UV luminosity is
discernible within our combined $z\geq 9$ sample, which again is consistent
with studies at lower redshift. Together, these results indicate that the
physical conditions governing nebular emission and its coupling to the UV
continuum emission properties and the ionising photon production efficiency in
galaxies are in place very early ($z \simeq 9-11$) on during the epoch of
reionisation and consistent with a continuation of trends already established
at $z \sim 6-9$.
\end{abstract}




\section{Introduction}
Ly$\alpha$ line emission from young, star-forming galaxies has long been a
powerful tracer of the early Universe, providing some of the earliest
spectroscopic confirmations of galaxies at high redshift
\citep[e.g.,][]{Partridge1967,Hu1998,Ouchi2008,Stark2010}. At early cosmic
times, Ly$\alpha$ efficiently traces star formation and the surrounding
intergalactic medium (IGM), owing to its high intrinsic luminosity and resonant
nature. However, during the Cosmic Dawn epoch ($z > 6$), the increasing neutral
hydrogen fraction of the IGM resonantly scatters Ly$\alpha$ photons, causing a
sharp decline in their detectability at fixed UV luminosity
\citep[e.g.,][]{Pentericci2011,Finkelstein2013,Treu2013,Pentericci2014,Caruana2014,Tilvi2014,Vanzella2014}.
This attenuation has traditionally made it extremely challenging to
spectroscopically confirm galaxies deemed to reside in the reionization era
based on their broadband rest-frame UV properties from deep {\it Hubble Space
Telescope} ({\it HST}) imaging \citep[e.g.,][]{Stanway2005, Bouwens2009,
McLure2010}. Detectable Ly$\alpha$ emission at these redshifts is typically
confined to rare, extreme systems capable of ionising large local bubbles that
allow Ly$\alpha$ photons to escape via resonant scattering
\citep[e.g.,][]{Oesch2015,Zitrin2015,Stark2017}.

Due to these limitations, attention increasingly shifted toward rest-frame
optical nebular emission lines, in particular H$\alpha$, H$\beta$, and
[O\textsc{iii}]$\lambda\lambda4959, 5007$, which are unaffected by the neutral
IGM and provide more reliable probes of early galaxy populations in the
reionization era
\citep[e.g.,][]{Labbe2013,Smit2015,DeBarros2019,Endsley2021,Endsley2023a}. The
strengths of these lines, commonly quantified via their rest-frame equivalent
widths (EWs), are sensitive to fundamental galaxy properties including
star-formation rate, stellar age, metallicity, and the hardness of the ionising
radiation field. As a result, a growing number of studies have focused on
characterising how nebular emission-line EWs scale with UV luminosity, stellar
mass, and UV continuum slope ($\beta$), and how these relations evolve with
redshift
\citep[e.g.,][]{Schenker2013,Khostovan2016,Faisst2016,Reddy2018,Tang2019,Endsley2021,Topping2022,Simmonds2024a,Boyett2024,Begley2025}.
These studies generally find that galaxies with lower masses, fainter UV
luminosities, and bluer UV slopes exhibit larger EWs, reflecting younger, more
metal-poor stellar populations with elevated specific star-formation rates.

Prior to the launch of the {\it James Webb Space Telescope} ({\it JWST}),
rest-frame optical emission line properties at  $z \gs 6$ were inferred from
broadband flux excesses in \textit{Spitzer}/IRAC photometry. Several studies
found that a significant fraction of galaxies at $z \simeq 7-8$ exhibit strong
H$\beta$+[O\textsc{iii}] emission, with median EWs of $\sim
600-700\,\si{\angstrom}$ \citep{Labbe2013,DeBarros2019,Endsley2021}, and that
$\sim 20$\% of those show extreme EWs exceeding $1000\,\si{\angstrom}$
\citep{Smit2015,Roberts-Borsani2016,Castellano2017}. These extreme emitters
were interpreted as rapidly assembling, metal-poor systems dominated by
short-lived O and B stars, capable of producing intense ionising radiation
while maintaining weak optical continua.

With {\it JWST}, direct spectroscopic constraints on rest-frame optical lines
are now possible at $z \gs 6$. Large surveys utilizing NIRCam grism
spectroscopy have identified hundreds of H$\beta$+[O\textsc{iii}] emitters at
$z \simeq 7-9$, finding median EWs consistent with pre-{\it JWST} IRAC-based
estimates \citep{Oesch2023,Meyer2024,Meyer2025}. Complementary broadband
studies using NIRCam imaging have extended these measurements to fainter
galaxies and revealed well-defined scaling relations between EW, UV luminosity,
and $\beta$ at $z \simeq 7-9$
\citep[e.g.,][]{PG2023,Rinaldi2023,Endsley2024,Boyett2024}.

A key motivation for characterising nebular emission at high redshift is its
close connection to the ionising photon production efficiency, $\xi_{\rm ion}$,
which sets the number of hydrogen-ionising photons produced per unit UV
luminosity. This quantity plays a central role in models of cosmic
reionization, linking observed galaxy populations to the ionising photon budget
of the early Universe. Recent {\it JWST} studies have shown that $\xi_{\rm
ion}$ correlates strongly with nebular EWs, UV luminosity, and UV slope, and
may evolve only weakly with redshift once these dependencies are accounted for
\citep[e.g.,][]{Tang2019,Endsley2023a,Simmonds2024a,Boyett2024,Begley2025}.
Establishing whether these scaling relations persist into the earliest epochs
is crucial for understanding the role of faint galaxies in driving
reionization.

Beyond $z \gs 9$, however, direct constraints on H$\beta$+[O\textsc{iii}]
emission remain sparse. Spectroscopic measurements are currently limited to a
small number of individual galaxies \citep[e.g.,][]{Hsiao2024b,
Alvaro-Marquez2025, Helton2025, Calabro2024, Heintz25, Harikane2026}, while
broadband studies using NIRCam become ineffective once the lines redshift
beyond the F444W  band at $z>9$. At the same time, galaxies at these redshifts
are on average expected to be increasingly metal-poor and may host very young
stellar populations, possibly dominated by Population~III stars with extreme
ionization parameters \citep[e.g.,][]{Inoue2011,Nakajima2022}. Extrapolating
trends observed at lower redshifts, such conditions might naively be expected
to produce very large nebular EWs. However, [O\textsc{iii}] emission is highly
sensitive to metallicity and oxygen abundance, and at sufficiently low
metallicities the [O\textsc{iii}]/H$\beta$ ratio may decline. In this regime,
the total H$\beta$+[O\textsc{iii}] EW could therefore flatten or even decrease
despite intense star formation \citep[e.g.,][]{Endsley2023b, Korber2025}. 

Indirect evidence for this scenario was reported by \citet{Trussler2024}, who
analysed a sample of $z\sim 10.5$ NIRCam-selected galaxies with non-negligible
Balmer breaks. Their best-fitting SEDs implied modest H$\beta$+[O\textsc{iii}]
EWs ($\sim 160\,\si{\angstrom}$), suggesting suppressed [O\textsc{iii}]
emission in very metal-poor systems. A strong inverse correlation between the
Balmer break strength and line EW was found, reinforcing the notion that the
youngest, most metal-deficient galaxies exhibit weaker rest-frame optical
emission lines. Recently, \citet{Harikane2026} reported a $z\sim11$ galaxy
showing a pronounced Balmer break together with weak rest-frame optical line
emission (H$\alpha$ and [O\textsc{iii}]$\lambda 5007$), providing further
evidence that some of the earliest galaxies may already host evolved stellar
populations and comparatively low nebular EWs. Determining whether the
H$\beta$+[O\textsc{iii}] EW distribution declines, flattens, or remains
elevated at $z \gs 9$ therefore provides a critical test of early chemical
enrichment, stellar population ages, and ionising conditions in the first
generations of galaxies.

In this paper, we exploit ultra-deep imaging from the MIRI Deep Imaging Survey
\citep[MIDIS;][]{Oestlin2025} to push the study of H$\beta$+[O\textsc{iii}]
line emitters to $z \gs 9$, using the F560W ($\sim 5.6\,{\rm \mu m}$) filter of
MIRI \citep{Rieke2015,Wright2015,Wright2023}. At these redshifts, both H$\beta$
and [O\textsc{iii}] fall within this bandpass, enabling us to probe a critical
yet largely unexplored epoch in the evolution of rest-frame optical line
emission. By combining MIRI observations with deep {\it HST} and NIRCam
imaging, we characterise the EW distribution, scaling relations, and ionising
efficiencies of faint galaxies deep into the epoch of reionization. Throughout
this paper, we adopt a $\Lambda$CDM cosmology with $H_0 = 70\,{\rm
km\,s^{-1}\,Mpc^{-1}}$, $\Omega_{\rm m} = 0.3$, and $\Omega_{\Lambda} = 0.7$.
Also, unless otherwise stated, we adopt in this paper the AB magnitude system
\citep{Oke1983} and assume a \citet{Chabrier2003} initial mass function (IMF).

\section{The MIDIS Survey}\label{section:MIDIS-survey}
\subsection{The data}\label{subsection: data}
The MIRI Deep Imaging Survey (MIDIS) is a deep MIRI survey of the Hubble Ultra
Deep Field \citep[HUDF;][]{Beckwith2006} undertaken by the MIRI European
Consortium GTO program (proposal ID 1283, PI: G.~Östlin). The program was
intended to integrate for 60 hours in the MIRI/F560W filter. However, due to a
safety-shutdown of {\it JWST} in December 2022, $41.3$ hours of on-source time
was obtained, initially, with an additional $\sim 10$ hours obtained a year
later, along with $\sim 10$ hours of integration in the F1000W band
\citep[see][]{PG2024}. An extensive and detailed description of the MIDIS
survey, including observations and data reduction, is given in
\citet{Oestlin2025}. The survey reaches a limiting magnitude in F560W of
$28.59\,{\rm mag}$ (AB, 5-$\sigma$ point-source sensitivity), and covers a
total area of $4.7\,{\rm sq.~arcmin}$.

The HUDF boasts one of the richest and deepest multiwavelength ancillary
datasets of all the extragalactic fields. In this paper, we make use of the
publicly available {\it JWST} and {\it HST} imaging in the HUDF. This includes
NIRCam imaging obtained by the JADES programs \citep[proposal ID
1180;][]{Rieke2023, Eisenstein2023}, as well as {\it HST}/ACS+WFC3 imaging
\citep{Illingworth2013}. 
the \texttt{Grizli} pipeline \citep{Brammer2021,Brammer2022}, and drizzled to a
resolution of $0.04\,{\rm \arcsecs/pixel}$ (see \citet{Oestlin2025} for further
details).

\subsection{Source extraction, photometry and photometric redshifts}\label{section:SED-fitting}
A source catalog was created using the {\tt The Farmer}
code\footnote{\url{https://github.com/astroweaver/the_farmer}}
\citep{Weaver2019, Weaver2022}, a tool that utilises {\tt SEP} (SExtractor for
Photometry)\footnote{\url{https://github.com/kbarbary/sep}} for the initial
source detection. A detailed description of the catalog is given in
\citet{Gillman2025}. The MIRI/F560W image served as the detection image, which
was combined with the inverse variance image as a weight map. Based on
extensive testing, we adopted the following detection parameters:
\texttt{THRESH}\,=\,3, \texttt{MIN\_AREA}\,=\,3 and
\texttt{FILTER\_KERNEL}\,=\,1.5\_3$\times$3.conv.  {\tt The Farmer} employs
profile-fitting to estimate the total photometry for the extracted sources,
eliminating the need for explicitly applying an aperture corrections.
Photometry in other bands (NIRCam + {\it HST}) is derived by forcing the fitted
model with only  the overall brightness as a free parameter
\citep[e.g.,][]{Weaver2022}.

With our multi-wavelength catalog, we fit spectral energy distributions (SEDs)
and derive photometric redshifts for all our sources, using {\tt
EAzY-py}\footnote{\url{https://github.com/gbrammer/eazy-py}}, which is an
updated version of the photometric redshift code {\tt EAzY}
\citep{Brammer2008}. {\tt EAzY-py} utilises an $\chi^2$-minimisation procedure
in which linear combinations of template SEDs are tested at different redshifts
to find an optimal fit to the observed fluxes. We use 13 templates from the
Flexible Stellar Populations Synthesis code \citet[FSPS;][]{Conroy2010}, which
cover a wide range of galaxy types and utilise a \cite{Chabrier2003} initial
mass function (IMF) and a \cite{Calzetti1994} dust attenuation law while
assuming solar metallicity. An advantage of these templates is that they
include emission lines, such that a narrowband excess can provide a relatively
tight constraint on the redshift. 

\section{Strong H$\beta$+[O\textsc{iii}] excess sources in the MIRI F560W band}
\subsection{Selection}\label{section:selection}
To identify sources exhibiting H$\beta$+[O\textsc{iii}] excess in the MIRI
F560W band, we applied a sequence of selection criteria to the full catalog.
The first criterion ensured that the observed photometry in at least one of the
NIRCam bands near F560W was consistent with the modeled continuum of the source
SED in the same band. Specifically, we required that the absolute difference
between the observed and model-predicted magnitudes satisfy $|m_{\rm X, obs} -
m_{\rm X, SED}| \leq 3\times \sigma_{\rm X, obs}$, where X corresponds to one
of the F430M, F444W, F460M, or F480M bands, and $\sigma_{\rm X, obs}$ is the
photometric uncertainty in that band. A total of 3817 sources passed this
initial selection.

\begin{figure*}
    \centering
    \includegraphics[width=\linewidth]{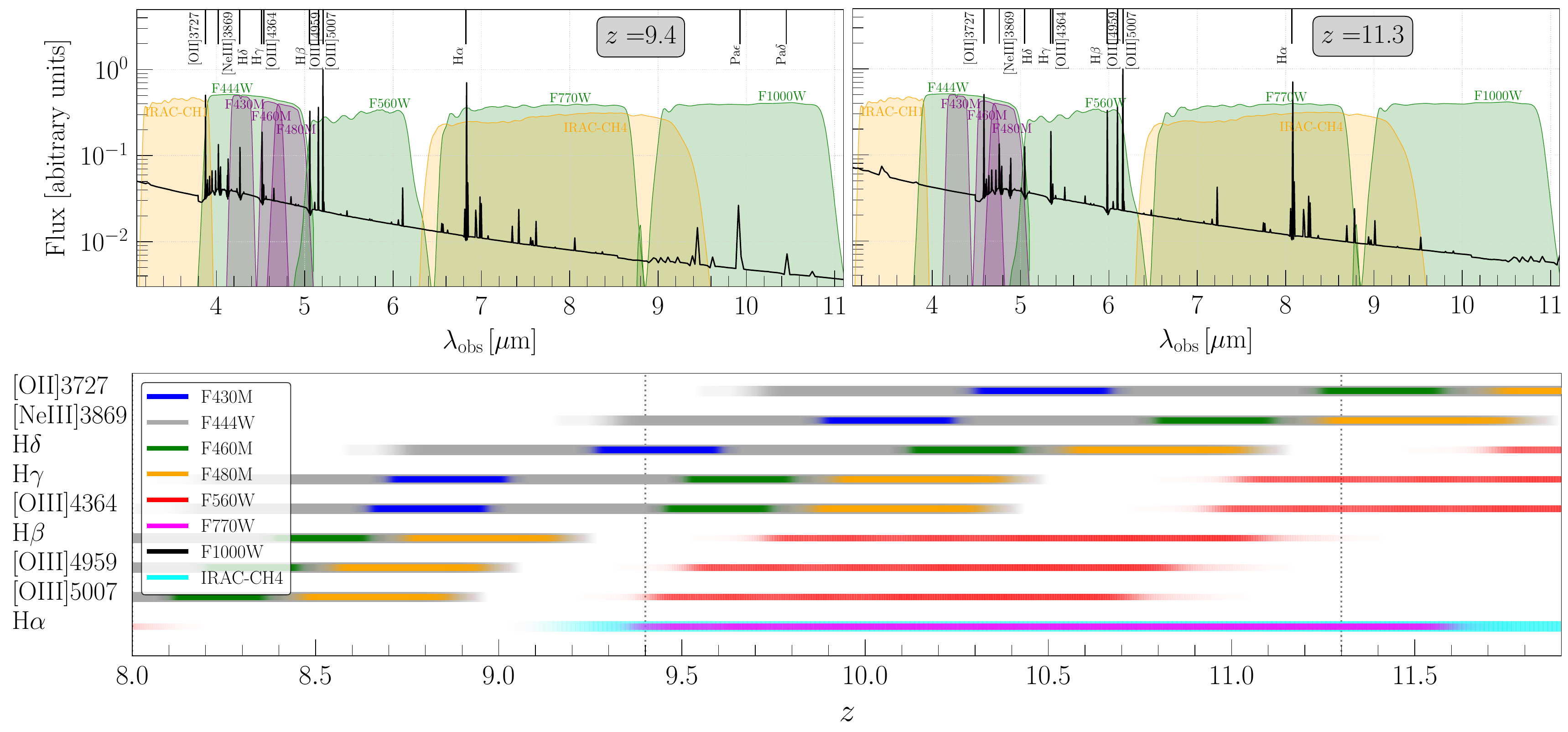}
    \caption{{\bf Top panels:} a model galaxy SED redshifted to $z=9.4$ (left
    panel) and $z=11.3$ (right panel), with the NIRCam/F430M, F444W, F460M,
    F480M and MIRI/F560W passbands overlaid. Also shown are the {\it Spitzer}/IRAC
    $3.6\,{\rm \mu m}$ and $8.0\,{\rm \mu m}$ passbands. The galaxy SED is taken
    from the LYCAN project \citep{Zackrisson2017}, based on the galaxy simulations
    by \citet{Gnedin2014}, and is used here solely to illustrate how the
    H$\beta$+[O\textsc{iii}] emission lines enter and exit the F560W filter at
    $z=9.4$ and $z=11.3$, respectively. The SED includes nebular emission lines,
    with key features such as  H$\alpha$, H$\beta$, H$\gamma$ and
    [O\textsc{ii}]\AA3728 labeled. {\bf Bottom panel:} The redshift evolution of
    the most prominent optical emission lines across the range $z=8-11.8$,
    indicating how they fall into the different passbands. The filters are
    color-coded as shown in the legend; for each filter, the intensity of the color
    reflects the transmission function of the passband.}
    \label{fig:illustration-of-selection-I}
\end{figure*}

These NIRCam bands are located just blueward of F560W and provide a reliable
estimate of the underlying continuum, provided no strong emission lines fall
within them. This criterion is similar to the one used in \citet{Rinaldi2023}.
As shown in Fig.\,\ref{fig:illustration-of-selection-I}, the main emission
lines that can affect these bands over the redshift range $z\sim 9.4-11.3$, are
[O{\sc ii}]$\lambda\lambda3727,3730$,  H$\delta\,\lambda4103$,
H$\gamma\,\lambda4342$ and [O\textsc{iii}]$\lambda 4364$. However, observations
of high-$z$ galaxies \cite[e.g.,][]{Schaerer2022,Sanders2023a, Sanders2023b}
indicate that these lines are significantly weaker, typically only a few
percent of the H$\beta$ and [O\textsc{iii}]$\lambda\lambda$4959, 5007 lines,
and therefore unlikely to significantly contaminate the continuum fluxes in
these NIRCam bands. This allows us to treat them as clean baseline measurements
of the continuum just blueward of F560W. The second criterion required an
excess in the F560W band relative to the NIRCam continuum bands. Specifically,
we selected sources with $m_{\rm F560W, obs} - m_{\rm X, obs} \leq -0.2$. This
corresponds to a flux excess of 20\% or more, which is similar to the excess
criterion adopted by \citet{Rinaldi2023} in their selection of $z\simeq 7-8$
H$\beta$+[O\textsc{iii}] emitters. Our approach is arguably more conservative,
since we require the flux excess to be with respect to the observed fluxes in
the F430M, F444W, F460M, or F480M  NIRCam bands, while \citet{Rinaldi2023} did
their selection based on a comparison with the F460M magnitude of their best
SED fit.

Applying our selection yielded 3107 sources showing a significant F560W excess.
As a third step, we examined the SEDs and photometric redshift probability
distribution functions, $p(z)$, generated using {\tt EAzY-py}, to select
galaxies within the redshift range where both H$\beta$ and [O\textsc{iii}] are
expected to fall within the F560W band. The [O\textsc{iii}] lines enter the
F560W passband for $z=9.1-11.3$, while H$\beta$ enters for $z=9.4-11.7$. We
restricted our sample to the overlapping redshift interval $z = 9.4-11.3$,
ensuring both lines contribute to the observed excess. We retained sources for
which the median of $p(z)$ lies within this range, resulting in 69 candidates.
We then applied an additional quality cut, requiring the reduced  $\chi^2$ of
the SED fit to be less than 3 to ensure good model agreement. Finally, we
visually inspected all remaining sources to exclude objects located near
diffraction spikes from bright stars or otherwise deemed spurious. This
multi-step selection process yielded a final sample of three robust
H$\beta$+[O\textsc{iii}] excess candidate sources, summarized in Table
\ref{tab:candidate-list}.
\begin{figure}
    \centering
    \includegraphics[width=0.95\linewidth]{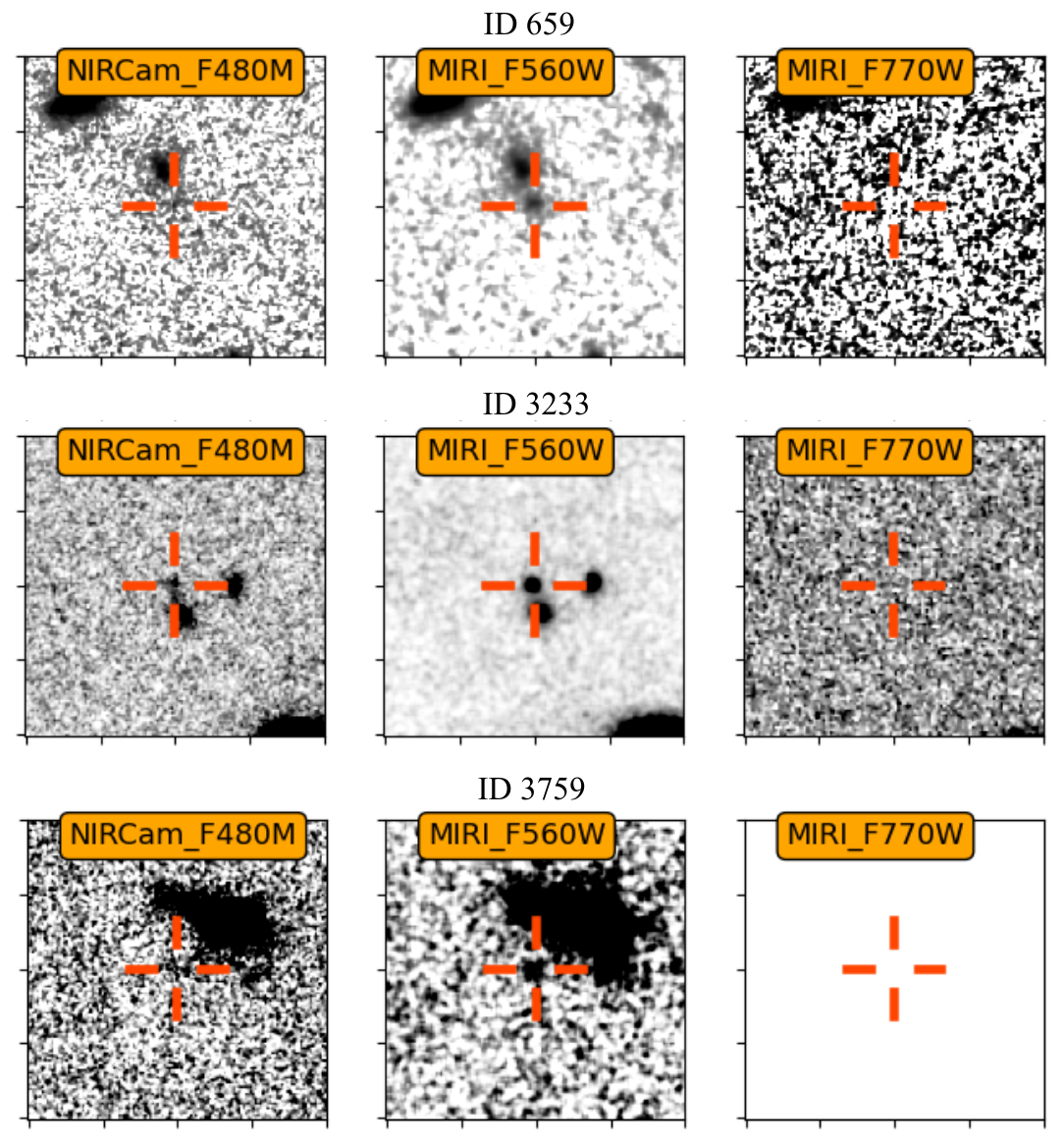}
    \caption{Postage stamps ($5\arcsecs \times 5\arcsecs$) images
    (NIRCam/F480M, MIRI/F550W, and MIRI/F770W) centered on our robust
    H$\beta$+[O\textsc{iii}] excess candidates. The images are displayed on the
    same flux scale.}
    \label{fig:postamps}
\end{figure}

\begin{figure*}
    \centering
    \includegraphics[width=0.88\linewidth]{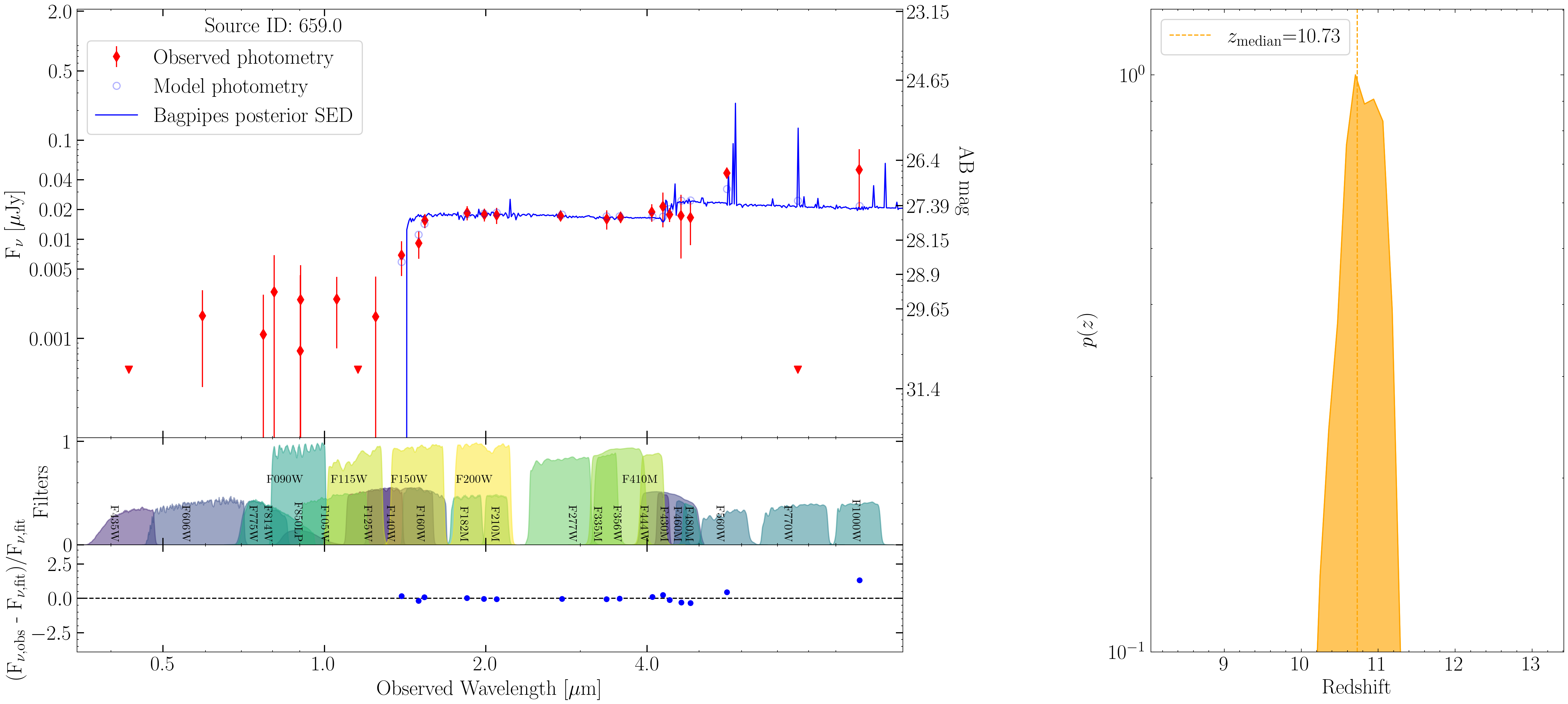}
    \includegraphics[width=0.88\linewidth]{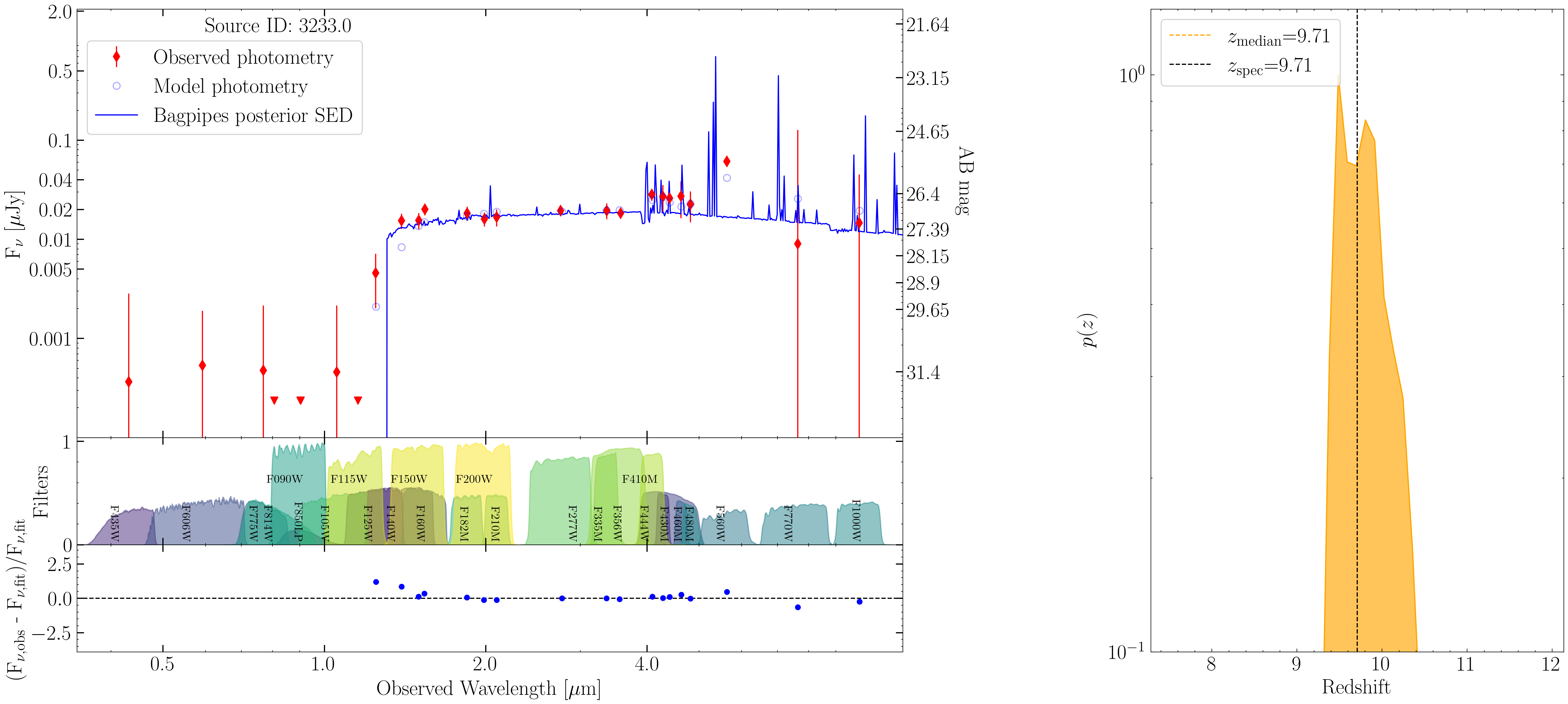}
    \includegraphics[width=0.88\linewidth]{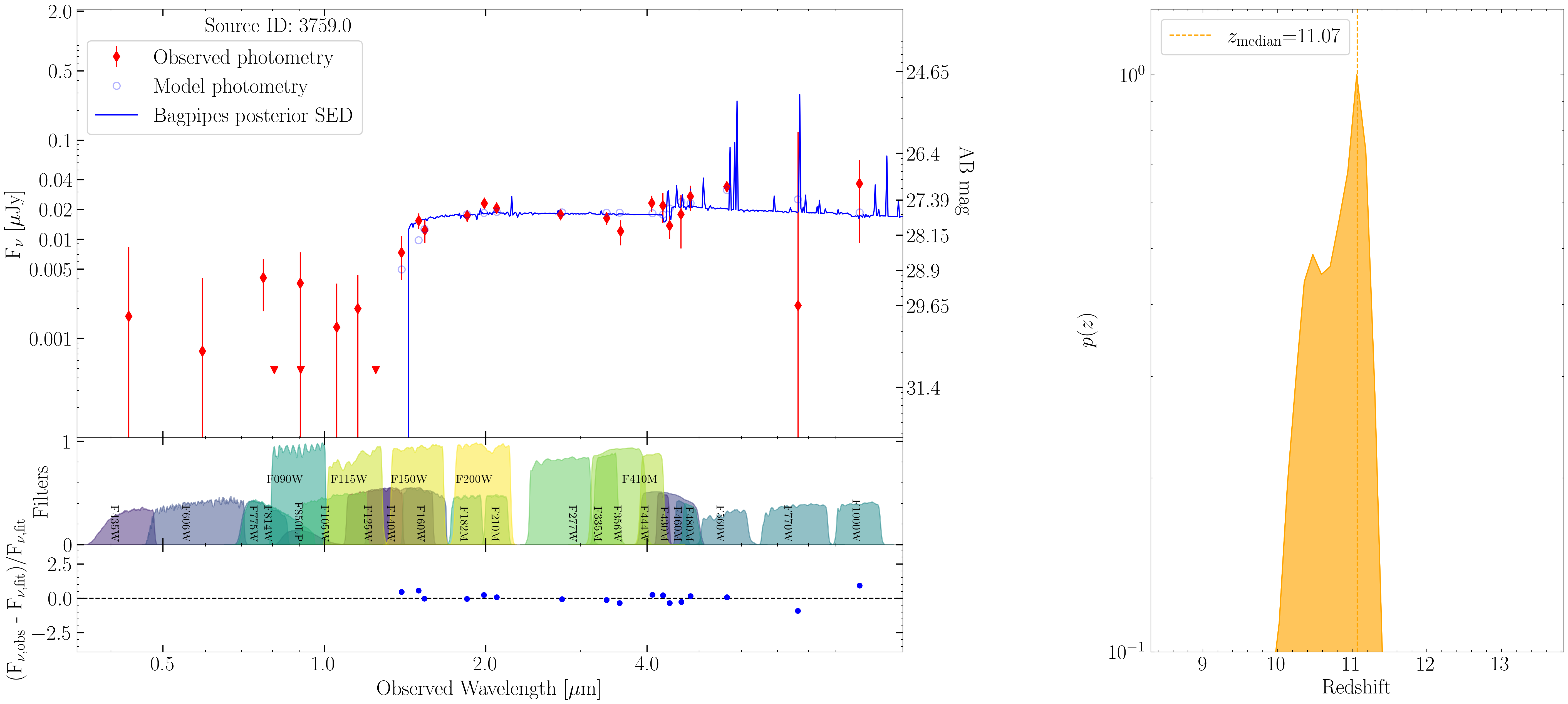}
    \caption{The observed broad-band photometry (red symbols) of galaxies
        ID\,659 (top), 3233 (middle) and 3759 (bottom), and the {\tt Bagpipes}
        best-fit SED models (blue curve) to the data. Positive and negative
        fluxes are shown as red filled circles and downward-pointing triangles,
        respectively. Also shown are the filter transmission functions of the
        corresponding {\it HST} and {\it JWST} filters, with the normalised
        residuals between the observed and best-fit fluxes in each filter shown
        in the bottom panel. The panel on the right shows the redshift
        probability distribution function, $p(z)$, derived from {\tt EAzY-py}.}
    \label{fig:659-example}
\end{figure*}

In Fig.\,\ref{fig:postamps} we show postage stamp images ($5\arcsecs\times
5\arcsecs$ in size) in the NIRCam/F480M, MIRI/F560W, and MIRI/F770W bands for
our robust candidates. As expected, all show a significant brightening in the
F560W band. In Fig.\,\ref{fig:659-example} (right panels) is shown the
posterior probability distribution function for the photometric redshift,
$p(z)$, based on the  {\tt EAzY-py} SED fit to our three robust candidates.  

In the redshift interval considered ($z=9.4-11.3$), the H$\alpha$ line falls
within the MIRI/F770W band (Fig.\,\ref{fig:illustration-of-selection-I}). The
fact that none of our H$\beta$+[O\textsc{iii}] emitters exhibit a significant
MIRI/F770W flux excess can be attributed to the shallowness of the observations
in this band compared to MIRI/F560W. 

Finally, we cross-matched the positions of our candidates against spectroscopic
redshifts available in the DAWN {\it JWST}
Archive\footnote{https://dawn-cph.github.io/dja/} \citep{Heintz25}. This
resulted in the spectroscopic redshift confirmation ($z_{\rm spec} = 9.721\pm
0.001$; \citet{Hainline2024}) of one of our sources (ID 3233: $z_{\rm phot} =
9.7^{+0.2}_{-0.1}$, see Table \ref{tab:candidate-list}).  

\begin{table*}
\caption[\protect]{Coordinates, photometric (and spectroscopic where available)
    redshifts, rest-frame UV ($1500\,$\AA) absolute magnitudes, UV spectral
    slopes, stellar masses and star-formation rates estimated for the three
    sources identified as strong H$\beta$+[O\textsc{iii}] line emitters in the
    MIDIS F560W image of XDF (\S\ref{section:selection}). The last two column
    lists the H$\beta$+[O\textsc{iii}] rest-frame equivalent widths and line
    luminosities of the sources, as derived in \S\ref{subsection:EW-values}.
    The upper and lower errors on the physical quantities are the 84th and 16th
    percentiles, respectively. 
}
\label{tab:candidate-list}
\centering
\setlength{\tabcolsep}{3pt}
\renewcommand{\arraystretch}{1.1}
\makebox[\textwidth][c]{%
\begin{tabular}{l c c c c c c c c c}
\hline
\hline
ID & RA & Dec & $z_{\rm phot}$ & $M_{\rm UV}$ & $\beta$ & $\log(M_{\star}/{\rm M_{\odot}})$ & SFR & ${\rm EW}^{\rm H\beta+[O\textsc{iii}]}_{\rm rest}$ & $L_{\rm H\beta+[O\textsc{iii}]}$ \\
 & hh:mm:ss.s & dd:mm:ss.s & & & & [dex] & [${\rm M_{\odot}\,yr^{-1}}$] & [\AA] & [$10^{42}\,{\rm erg\,s^{-1}}$] \\
\hline
659 &  03:32:35.2 & $-$27:47:38.7   &      $10.7^{+0.1}_{-0.1}$ & $-19.3^{+0.1}_{-0.1}$ & $-2.0^{+0.2}_{-0.2}$ & $8.0^{+0.4}_{-0.3}$ & $1.1^{+2.0}_{-0.6}$ & $1269^{+257}_{-259}$ & $4.0^{+1.0}_{-1.0}$ \\
3233$^{\dagger}$ & 03:32:42.1 & $-$27:46:50.3   & $9.7^{+0.2}_{-0.1}$ & $-19.2^{+0.1}_{-0.1}$ & $-1.8^{+0.2}_{-0.2}$ & $8.2^{+0.4}_{-0.5}$ & $1.9^{+1.9}_{-1.4}$ & $1307^{+217}_{-219}$ & $4.5^{+1.0}_{-1.0}$ \\
3759 & 03:32:43.7 & $-$27:46:47.8   &     $11.1^{+0.1}_{-0.7}$ & $-19.4^{+0.2}_{-0.1}$ &  $-2.1^{+0.2}_{-0.2}$ & $8.4^{+0.2}_{-0.2}$ & $2.7^{+0.7}_{-1.2}$ & $608^{+91}_{-92}$ & $1.8^{+0.5}_{-0.5}$ \\
\hline
\end{tabular}%
}
\vspace{2pt}
{\raggedright\footnotesize
$^{\dagger}$ Has a spectroscopically measured redshift of $z=9.721\pm 0.001$.\par
}
\end{table*}

\subsection{Physical properties}\label{subsection:physical_properties}
While the photometric redshifts for our sample were derived with {\tt EAzY-py},
we used the {\tt Bagpipes} (Bayesian Analysis of Galaxies for Physical
Inference and Parameter EStimations) SED fitting code
\citep{Carnall2018,Carnall2019} to derive their physical properties, e.g.,
stellar masses ($M_\star$), star-formation rates (SFRs), UV luminosities
($M_{\rm UV}$), and UV continuum slopes ($\beta$) for all three galaxies
(Table\,\ref{tab:candidate-list}).  

For the fitting, we adopted a single-component, exponentially declining
(“$\tau$-model”) star-formation history (SFH) with an optional additional young
component to capture recent small bursts of star-formation. The base SFH
component was parameterized by the stellar population age $t_{\rm{age}}$ and
the e-folding time ($\tau$), both varied within wide uniform priors:
$t_{\rm{age}}\in[0.01,0.3]\,\rm{Gyr}$ and $\tau\in[0.3,5.0]\,\rm{Gyr}$. The
total stellar mass formed was allowed to vary within $\log(M_\star/M_\odot) \in
[1, 10]$, and the metallicity within $Z/Z_\odot \in [0.02, 1.0]$. The young
component was parameterised with $t_{\mathrm{age}} \in [0.01, 0.15]$\,Gyr,
$\tau \in [0.05, 0.5]$\,Gyr, and $\log(M_\star/M_\odot) \in [1, 8]$.

Nebular line and continuum emission was included via the internal {\tt
Bagpipes} implementation, with the ionization parameter allowed to vary
uniformly within $\log U\in[-4.0,-1.0]$, and the escape fraction limited to
$f_{\rm esc}\in[0,0.2]$. Dust attenuation was modelled with the
\citet{Calzetti1994} law and a uniform prior on $A_V\in[0,0.3]$. A Gaussian
prior on the redshift was adopted, centred on the photometric estimate from
{\tt EAzY-py} and with a width corresponding to its associated uncertainty.
This approach incorporates the available photometric information while still
allowing the fit to explore the redshift range $z \in [8,12]$. For the
spectroscopically confirmed galaxy (ID~3233), the redshift was fixed to
$z_{\mathrm{spec}} = 9.721$ with a narrow dispersion ($\sigma_z = 0.001$) to
anchor the fit at the measured value.  The resulting best-fit SEDs from {\tt
Bagpipes} are shown in blue in Fig.\,\ref{fig:659-example}.

From the best-fit SED and 5000 samples of the posterior distribution, we derive
the physical properties and their associated 16th and 84th percentile
uncertainties (Table \ref{tab:candidate-list}). UV luminosities ($M_{\rm UV}$)
are not directly reported by {\tt Bagpipes} and were therefore measured from
the best-fit SED model. The UV luminosity were computed with a
$100\,\si{\angstrom}$ wide top-hat filter centred at a rest-frame wavelength of
$1500\,\si{\angstrom}$. We measured $\beta$ by fitting a power law ($F_\lambda
\propto \lambda^{\beta}$) to the stellar continuum (i.e., without nebular
emission) of the best-fit model in the rest-frame wavelength range
$1250-2600\,\si{\angstrom}$, excluding regions contaminated by strong emission
lines \citep[e.g.,][]{Calzetti1994}. In other works, $\beta$ is measured
directly from low-resolution ($R\sim 30-300$) spectroscopy
\citep[e.g.,][]{Dottorini2025}, i.e., the total (stellar + nebular). If young
populations dominate, the (stellar + nebular) continuum slope tends to be less
steep than the value for stellar only.

\subsection{F560W flux excess, rest-frame H$\beta$+[O\textsc{iii}] line equivalent widths and luminosities}\label{subsection:EW-values}
To quantify the strength of the H$\beta$+[O\,\textsc{iii}] emission feature, we
measured the flux excess in the MIRI/F560W band relative to the underlying
continuum. Accurate continuum determination is essential for deriving reliable
EWs, and several approaches have been adopted in the literature. This includes
performing SED fits that exclude filters contaminated by strong emission lines
and using the model-predicted continuum flux as a reference
\citep[e.g.,][]{marmol-queralto-2016, Smit2015}, or using a neighbouring,
line-free filter such as NIRCam/F460M or NIRCam/F480M as a proxy for the
continuum \citep[e.g.,][]{Rinaldi2023,Korber2025}. 

The observed flux excess in the F560W band was computed as  
\begin{equation}
\Delta m = m_{\mathrm{F560W,obs}} - m_{\mathrm{F560W,cont}},
\end{equation}
where $m_{\mathrm{F560W,obs}}$ is the observed magnitude and
$m_{\mathrm{F560W,cont}}$ is the magnitude of the continuum predicted by the
{\tt Bagpipes} model. The rest-frame equivalent width of the
H$\beta$+[O\,\textsc{iii}] feature is then given by  
\begin{equation}
{\rm EW_{\rm rest}^{\rm H\beta+[O\textsc{iii}]}} = \frac{W_{\mathrm{rec}}}{1+z} \left( 10^{-0.4\,\Delta m} - 1 \right),
\end{equation}
where $W_{\mathrm{rec}}$ is the rectangularized width of the F560W passband
\citep[e.g.,][]{marmol-queralto-2016}.  Uncertainties on the reported EWs were
estimated in the same way as the for the physical parameters, i.e., as the 16th
and 84th percentile values obtained from 5000 samples of the posterior {\tt
Bagpipes} distribution and  the photometric errors on the F560W flux. We note,
that the adopted flux excess criterion of $\Delta m \leq -0.2$ in F560W,
corresponds to a minimum rest-frame EW of $\simeq 66\,\si{\angstrom}$ at
$z=9.4$ and $\simeq 56\,\si{\angstrom}$ at $z=11.3$. 

In our analysis, we modelled the F560W band-averaged continuum using the {\tt
Bagpipes} posterior SED samples obtained from the SED fitting
(\S\ref{section:SED-fitting}). {\tt Bagpipes} allows us to generate both the
full model spectrum, including stellar, nebular continuum, and line emission,
and a purely stellar continuum spectrum by excluding nebular emission. This
enables us to estimate EWs relative to either {\it i)} the total continuum
(stellar + nebular) or {\it ii)} the stellar continuum alone. The latter
isolates the strength of the emission lines with respect to the underlying
stellar population and provides a more direct probe of the specific
star-formation rate. The nebular continuum can contribute significantly (up to
$20-40\%$) to the total rest-frame optical continuum in strongly ionised
systems \citep[e.g., ][]{Katz2025}. Some photometric studies report EWs
relative to the total continuum \citep[e.g.,][]{Labbe2013, Smit2015,
Endsley2021}, while other works correct for nebular continuum when comparing to
models \citep[e.g.,][]{Tang2019}. Here we provide EW values relative to the
total (stellar+nebular) continuum (see Table \ref{tab:candidate-list}). We note
that deriving EWs with respect to the stellar continuum instead of the total
continuum yields values that are $\sim 4-26\,\%$ higher. We find rest-frame
H$\beta$+[O\textsc{iii}] EWs ranging from $608\,\si{\angstrom}$ to
$1307\,\si{\angstrom}$, with a median value of $1260\,\si{\angstrom}$ and a
median absolute deviation (MAD) of $37\,\si{\angstrom}$. 

The H$\beta$+[O\textsc{iii}] line luminosites ($L_{\rm
H\beta+[O\textsc{iii}]}$) were derived from the measured F560W flux excess
relative to the continuum predicted by the {\tt Bagpipes}. Specifically, the
line flux is calculated as the difference between the measured F560W flux and
the total (stellar+nebular) F560W bandpass-averaged model continuum, and the
line luminosity is subsequently calculated using $L_{\rm
H\beta+[O\textsc{iii}]} = 4\pi D_{\rm L}^2 F_{\rm F560W}$, where $D_{\rm L}$ is
the luminosity distance. Uncertainties (16th and 84th percentiles) on the line
luminosities (Table \ref{tab:candidate-list}) account for both the photometric
measurement error in F560W and the the Monte Carlo sampling of 5000 SED
realizations from {\tt Bagpipes}. 

\section{Literature Sample of H$\beta$+[O\textsc{iii}] Emitters at $z \gs 9$}\label{section:literature-sample}
To place our MIDIS sample in a broader context, we compile all currently
available measurements of H$\beta$+[O\textsc{iii}] emission from galaxies at $z
\gs 9$ reported in the literature (Table \ref{tab:lit_z9_ew}). This compilation
includes both spectroscopic and photometric measurements of ${\rm EW}_{\rm
rest}^{\rm H\beta+[O\textsc{iii}]}$, together with associated physical
properties such as $M_{\star}$, $M_{\rm UV}$, $\beta$, and $\xi_{\rm ion,0}$
where available. The literature sample spans a range of observational
approaches, including NIRSpec spectroscopy out to $z\simeq 9.5$ from the
PRImordial gas Mass AssembLy (PRIMAL) survey \citep{Heintz25}, MIRI
spectroscopy \citep{Hsiao2024b, Zavala2025,Helton2025, Alvaro-Marquez2025,
Alvarez-Marquez26, Harikane2026, Marques-Chaves26}, and broadband flux-excess
measurements based on MIRI imaging \citep{Crespo-Gomez26}. By assembling this
literature sample, we aim to assess how representative our MIDIS galaxies are
relative to previously reported $z\gs 9$ systems, and to explore trends in
${\rm EW}_{\rm rest}^{\rm H\beta+[O\textsc{iii}]}$, $M_{\rm UV}$, $M_{\star}$,
$\beta$, and $\xi_{\rm ion}$ at the highest redshifts currently accessible.

\begin{figure}
    \centering
    \includegraphics[width=\linewidth]{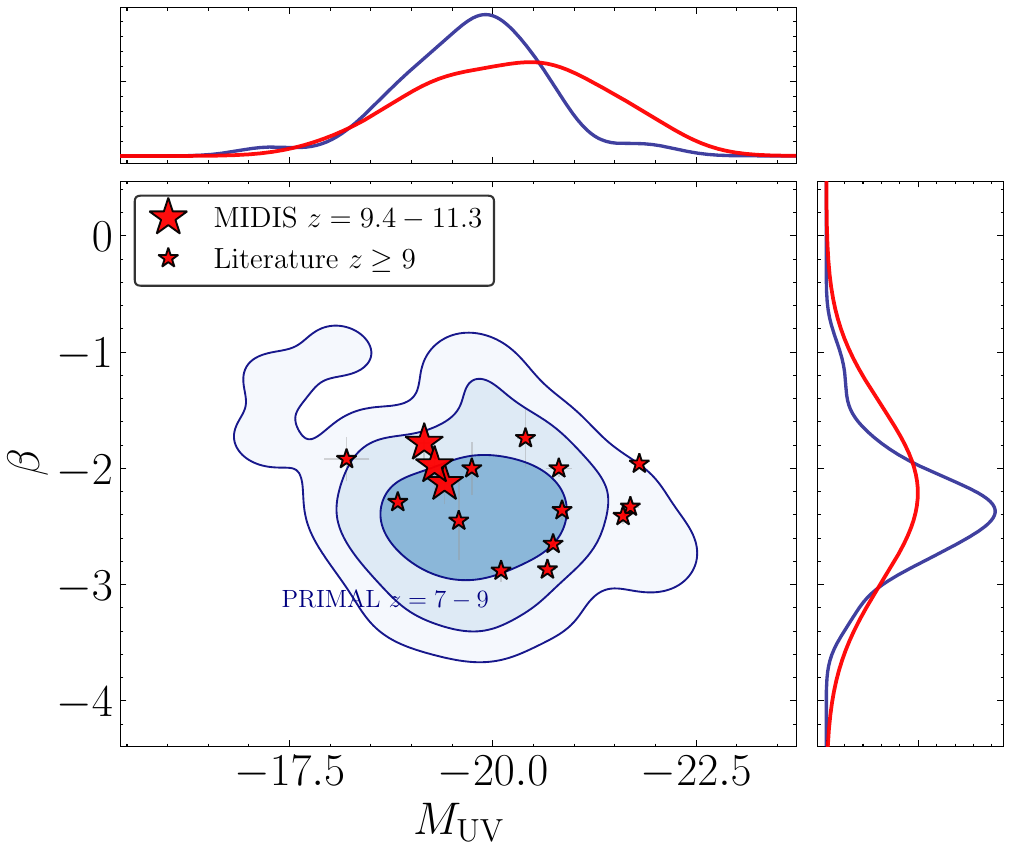}
    \caption{The distribution of our MIDIS $z=9.4-11.3$ sample and the $z\geq 9$ literature sample in the $\beta$-$M_{\rm UV}$ plane.
    For the literature sample, the $M_{\rm UV}$ values have been corrected for lensing where appropriate.
    For reference we also show the 2D KDE distribution of the $z=7-9$ sample from the PRIMAL survey \citep[blue contours;][]{Heintz25}. 
    The top and right panels show the 1D KDE distributions for $M_{\rm UV}$ and $\beta$, respectively for the combined $z\geq 9$ (MIDIS+literature)
    sample (red curve) and the $z=7-9$ PRIMAL sample (blue curve).
    }
    \label{fig:muv_beta}
\end{figure}

For most sources, we adopt the published ${\rm EW}_{\rm rest}^{\rm H\beta+[O\textsc{iii}]}$ 
values directly from the literature. However, for four sources (marked with daggers 
in Table \ref{tab:lit_z9_ew}), where no H$\beta$+[O\textsc{iii}] EWs estimates were provided, we derived 
the equivalent widths ourselves from the  available data as described below.\newline
\noindent \textbf{RXJ2129-11027 ($\boldsymbol{z} = 9.51$):}
\citet{Williams2023} detects H$\beta$ in this strongly lensed galaxy and
reports ${\rm EW}_{\rm rest}^{\rm H\beta} = 248\pm 35\,\si{\angstrom}$
\citep[see also][]{Langeroodi2023}. Adopting the \citet{Korber2025} $M_{\rm
UV}$-dependent relation for [O\textsc{iii}]$\lambda5007/{\rm H}\beta$ (see also
\S\ref{subsubsection:xi_ion}), and assuming the atomic doublet flux ratio
[O\textsc{iii}]$\lambda5007/$[O\textsc{iii}]$\lambda4959 = 2.98$\footnote{this
ratio is fixed by the Einstein $A$ coefficients of the O$^{++}$ ion and
therefore independent of nebular conditions \citep[e.g.,][]{Osterbrock2006}.},
the delensed $M_{\rm UV} = -18.7$ implies
[O\textsc{iii}]$\lambda\lambda4959,5007/{\rm H}\beta \simeq 7.93$. Combined
with the directly measured ${\rm EW}_{\rm rest}({\rm H}\beta) = 248 \pm
35\,\si{\angstrom}$, this yields ${\rm EW}_{\rm rest}({\rm H}\beta + [{\rm
O\textsc{iii}}]) \simeq 2215 \pm 313\,\si{\angstrom}$.
\newline 
\noindent \textbf{MACS0647--JD ($\boldsymbol{z} = 10.165$):}
\citet{Hsiao2024b} reports [O\textsc{iii}]$\lambda\lambda$4959,5007 and
H$\alpha$ line fluxes. Since H$\beta$ is not  directly detected in this
spectrum, we infer its flux from the measured H$\alpha$ line assuming Case B
recombination with $F_{\rm H\alpha}/F_{\rm H\beta} = 2.86$, appropriate for
$T_e \approx 10^4$ K and negligible dust attenuation. In order to derive ${\rm
EW}_{\rm rest}^{\rm H\beta+[O\textsc{iii}]}$, we estimate the continuum at
$\sim 5.6\,{\rm \mu m}$  (observed frame) from the two reddest broadband
photometry available \citep[F444W and F480M;][]{Hsiao2024a} using a weighted
average that assumes a flat spectrum. We find ${\rm EW}_{\rm rest}^{\rm
H\beta+[O\textsc{iii}]} \simeq 970\pm 200\,\si{\angstrom}$, where the error
includes the uncertainties in the measured line and broadband fluxes (amounting
to a mere 7\% error) as well as an assumed systematic error of 20\% due to the
uncertainty in the extrapolation of the continuum to $5.6\,{\rm \mu
m}$.
\newline
\noindent \textbf{CEERS2--588 ($\boldsymbol{z} = 11.04$):}
\citet{Harikane2026} reports upper limits on the [O\textsc{iii}]$\lambda$5007
and H$\alpha$ rest-frame equivalent widths. As for MACS0647$-$JD, we infer the
H$\beta$ flux from H$\alpha$ assuming $F_{\rm H\alpha}/F_{\rm H\beta} = 2.86$.
The [O\textsc{iii}]$\lambda$4959 line flux is derived adopting the atomic
doublet ratio $F_{\lambda5007}/F_{\lambda4959} \simeq 3$. Assuming the same
underlying continuum for H$\beta$ and [O\textsc{iii}], we can add the EWs in
proportion to the line fluxes. This yields ${\rm EW}^{\rm
H\beta+[O\textsc{iii}]}_{\rm rest} \ls 600\,\si{\angstrom}$.
\newline
\noindent \textbf{GLASS--z12 ($\boldsymbol{z} = 12.33$):}
All three lines (H$\beta$ and [O\textsc{iii}]$\lambda\lambda$4959,5007) are
directly measured, and the continuum at $\sim 6.6\,{\rm \mu m}$ is extrapolated
from the F444W photometry assuming flat $F_\nu$ \citep{Zavala2025}. We derive
${\rm EW}_{\rm rest}^{\rm H\beta+[O\textsc{iii}]} \simeq 1450\pm
312\,\si{\angstrom}$, where as for MACS0647$-$JD we have allowed for a 20\%
systematic uncertainty in the continuum extrapolation to $\sim 6.6\,{\rm \mu
m}$.
\newline
\noindent \textbf{UNCOVER--37126 ($\boldsymbol{z} = 10.255$):}
We derive  a $3\sigma$ upper limit on ${\rm EW}^{\rm
H\beta+[O\textsc{iii}]}_{\rm rest}$ by summing the reported individual EW upper
limits on H$\beta$ and [O\textsc{iii}]$\lambda$5007 \citep{Marques-Chaves26},
and accounting for the [O\textsc{iii}]$\lambda$4959 line using the atomic flux
ratio $F_{\lambda5007}/F_{\lambda4959} = 2.98$. This yields ${\rm EW}^{\rm
H\beta+[O\textsc{iii}]}_{\rm rest} < 300\,\si{\angstrom}$. 

\bigskip

From Fig.\,\ref{fig:muv_beta} we see that our MIDIS broadly occupy the same
part of the $\beta - M_{\rm UV}$ parameters space as the $z\geq 9$ literature
sources, although the former are on the faint-end in terms of $M_{\rm UV}$
(median $\langle M_{\rm UV}\rangle = -19.3\pm 0.1$ vs $\langle M_{\rm
UV}\rangle = -20.6\pm 0.4$ for the literature sample). This is expected, given
the deep MIRI selection. Overall, the $z \geq 9$ sources (MIDIS + literature)
lie in the same region in the $\beta - M_{\rm UV}$ plane as the $z=7-9$ PRIMAL
sources from \citet{Heintz25}, with a slight tendency to be more UV luminous,
which we attribute to selection bias towards more luminous systems at higher
redshifts.

\begin{table*}
\centering
\caption{Compilation of published ${\rm EW}^{\rm H\beta+[O\textsc{iii}]}_{\rm
rest}$ measurements for galaxies at $z\geq 9$ used in this work. Where
available, relevant physical properties, i.e., $M_{\rm UV}$, $\beta$, and
$M_{\star}$ are listed.}
\label{tab:lit_z9_ew}
\setlength{\tabcolsep}{4pt}
\renewcommand{\arraystretch}{1.1}
\resizebox{\textwidth}{!}{
\begin{tabular}{l c c c c c l}
\hline
ID & $z_{\rm spec}$ & $M_{\rm UV}$ & $\beta$ & $\log(M_\star/\Msolar)$ & ${\rm EW}_{\rm rest}^{\rm H\beta+[O\textsc{iii}]}$ &  Reference \\
 &  &  & & [dex] &  [\AA] &   \\
\hline
PRIMAL--1       & 9.05809 & $-19.58$          & $-2.45\pm 0.34$ & 7.20     & 1543                    & \citet{Heintz25} \\
PRIMAL--2       & 9.11154 & $-21.69$          & $-2.33\pm 0.06$ & 9.17     & 1679                    &  \citet{Heintz25} \\
PRIMAL--3       & 9.25102 & $-20.68$          & $\cdots$        & 7.98     & 18504                   &  \citet{Heintz25} \\
PRIMAL--4       & 9.31994 & $-21.80$          & $-1.96\pm 0.04$ & 9.20     & 360                     &  \citet{Heintz25} \\
PRIMAL--5       & 9.37997 & $-20.85$          & $-2.36\pm 0.08$ & $\cdots$ & 2619                    &  \citet{Heintz25} \\
PRIMAL--6       & 9.43617 & $-20.67$          & $-2.87\pm 0.02$ & 8.56     & 1347                    &  \citet{Heintz25} \\
PRIMAL--7       & 9.43826 & $-20.74$          & $-2.65\pm 0.02$ & 8.56     & 2060                    &  \citet{Heintz25} \\
PRIMAL--8       & 9.50948 & $-19.74$          & $-2.00\pm 0.23$ & $\cdots$ & 11947                   &  \citet{Heintz25} \\
RXJ2129--11027  & 9.51    & $-18.20$          & $-1.92\pm 0.19$ & 7.63     & $2215\pm 313$           &  \citet{Williams2023, Langeroodi2023} \\
UNCOVER--26185  & 10.054  & $-18.83\pm 0.07$  & $-2.29\pm 0.06$ & 8.23     & $467\pm 81$             &  \citet{Alvarez-Marquez26}\\
MACS0647--JD    & 10.165  & $-20.30$          & $\cdots$        & 8.10     & $970\pm 200^{\dagger}$  & \citet{Hsiao2024b} \\
UNCOVER--37126  & 10.255  & $-20.10\pm 0.10$  & $-2.88\pm 0.10$ & 7.77     & $<300^{\dagger}$        & \citet{Marques-Chaves26} \\
GNz11           & 10.63   & $-21.60\pm 0.04$  & $-2.41\pm 0.07$ & 9.18     & $926\pm 83$             &  \citet{Alvaro-Marquez2025, Crespo-Gomez26}\\
CEERS2--588     & 11.04   & $-20.40$          & $-1.74\pm 0.25$ & 9.00     & $<600^{\dagger}$        & \citet{Harikane2026} \\
GLASS--Z12      & 12.34   & $-20.50$          & $-2.39\pm 0.07$ & 8.91     & $1450\pm 312^{\dagger}$ & \citet{Zavala2025, Calabro2024} \\
JADES-GS-z14-0  & 14.1796 & $-20.81\pm0.16$   & $-2.00\pm 0.07$ & 8.72     & $714\pm 207$            & \citet{Helton2025} \\
\hline
\end{tabular}
}
\vspace{2pt}
\raggedright
\footnotesize
$^{\dagger}$ See \S\ref{section:literature-sample} for how ${\rm EW}^{\rm H\beta+[O\textsc{iii}]}_{\rm rest}$ was derived.
\end{table*}

\section{Results \& Discussion}
\subsection{The distribution of H$\beta$+[O\textsc{iii}] rest-frame EWs}\label{subsection:EW-hist}
\subsubsection{A first look at the H$\beta$+[O\textsc{iii}] distribution}\label{subsection:first-look}
\begin{figure}
    \centering
    \includegraphics[width=\linewidth]{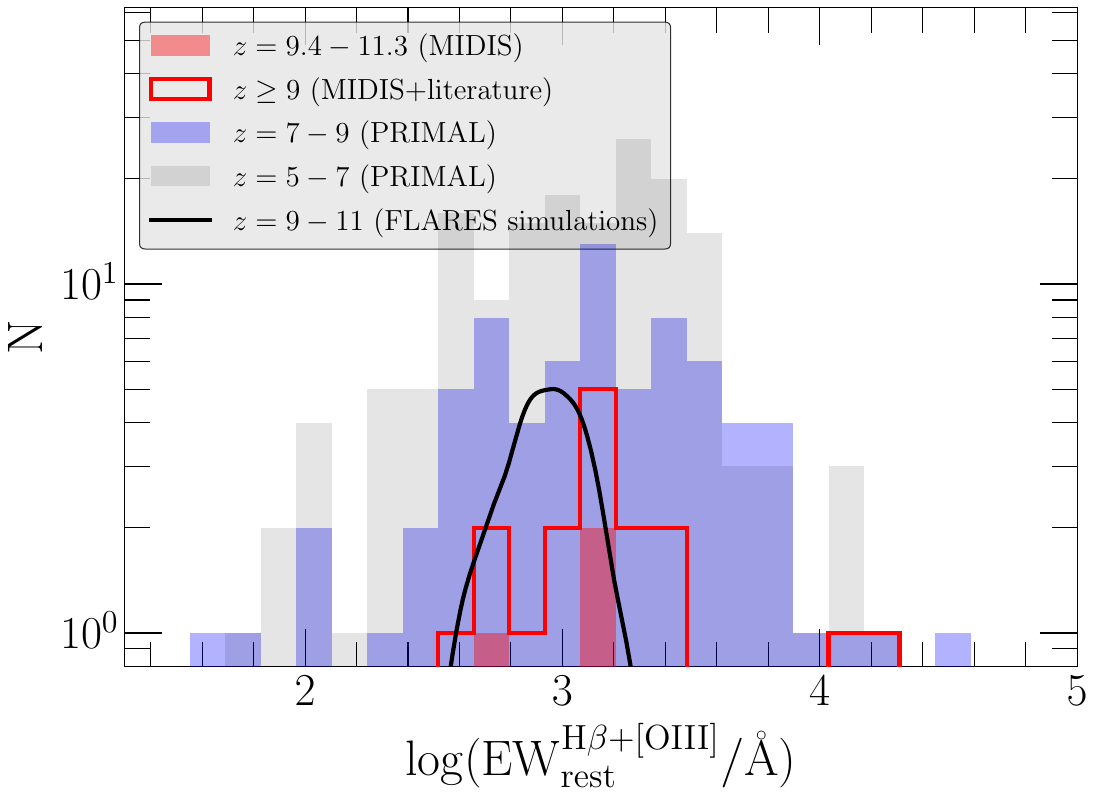}
    \caption{The H$\beta$+[O\textsc{iii}] EW distribution for our $z=9.4-11.3$
        MIDIS sample (red filled histogram) and the combined $z\geq 9$ MIDIS +
        literature sample (red open histogram). Shown as grey and blue filled
        histograms are the EW distributions for the $z=5-7$ and $7-9$
        subsamples of the PRIMAL survey \citep{Heintz25}. The black curve shows
        the kernel density estimator of the EW-distribution for the {\tt
        FLARES} simulations with the same observational selection imposed as
        for our MIDIS sample. The distribution has been normalised to the peak
        value ($N=5$) of the combined $z\geq 9$ distribution.}
    \label{fig:ew_hist}
\end{figure}

Fig.\,\ref{fig:ew_hist} shows the distribution of the derived
H$\beta$+[O\textsc{iii}] rest-frame EW values for our robust MIDIS sample of
H$\beta$+[O\textsc{iii}] line emitters (red filled histogram -- values listed
in Table \ref{tab:candidate-list}). We also include in Fig.\,\ref{fig:ew_hist}
the H$\beta$+[O\textsc{iii}] EW distribution for the 13 $z \geq 9$ galaxies
(excluding the two sources with upper limits on EW) gleaned from the literature
(red open histogram, see Table \ref{tab:lit_z9_ew}). 

With only 13 galaxies in the $z\geq 9$ literature sample, and three in our
MIDIS sample, it is difficult to robustly compare the two samples due to low
number statistics. Our MIDIS sample is selected from a small area and therefore
unlikely to be representative of the underlying distribution. Comparing the
${\rm EW}^{\rm H\beta+[O\textsc{iii}]}_{\rm rest}$ distributions, we find
nearly identical typical values, with a median $\log({\rm EW}^{\rm
H\beta+[O\textsc{iii}]}_{\rm rest}/\si{\angstrom}) = 3.10\pm 0.01$ ($\sim
1260^{+327}_{-259}\,\si{\angstrom}$) for the MIDIS sample and $3.16\pm 0.19$
($\sim 1497^{+794}_{-512}$) for the $z \gs 9$ literature sample, where the
scatter quoted is the median absolute deviation (MAD). The difference in the
medians is a mere $\sim 0.06\,{\rm dex}$ (corresponding to $\sim 2$\%
difference). A permutation test on the median yields no statistically
significant difference. Excluding the two highest-EW objects (PRIMAL--3 and
PRIMAL--8 with equivalent widths $\gs 10^4\,\si{\angstrom}$, see Table
\ref{tab:lit_z9_ew}) from the literature sample shifts its median to $\log({\rm
EW}^{\rm H\beta+[O\textsc{iii}]}_{\rm rest}/\si{\angstrom}) = 3.15\pm 0.17$,
decreasing the offset  further ($\sim 0.05\,{\rm dex}$). Given the
statistically indistinguishable medians and the small offset relative to the
intrinsic scatter, we cannot rule out that the two samples are probing the same
underlying population of $z\geq 9$ galaxies, despite their heterogeneous
selection. Combining the two samples, we show in Fig.\,\ref{fig:ew_hist} the
EW-distribution of the full (MIDIS+literature) $z\geq 9$ sample (red open
histogram). For this combined sample, we find a median $\log({\rm EW}^{\rm
H\beta+[O\textsc{iii}]}_{\rm rest}/\si{\angstrom}) = 3.12\pm 0.17$ ($\sim
1318^{+544}_{-385}\,\si{\angstrom}$).

For comparison with H$\beta$+[O\textsc{iii}] EW distributions lower redshifts,
we include H$\beta$+[O\textsc{iii}] measurements from the PRIMAL survey
\citep{Heintz25}, which presents NIRSpec spectroscopy for $\sim 600$ galaxies
at $z\gs 5.5$. We split the PRIMAL sample into EW distributions for galaxies in
the redshift ranges $z=5-7$ and $7-9$, while the $z \geq 9$ PRIMAL galaxies are
included in our overall $z \geq 9$ literature sample (see
\S\ref{section:literature-sample} and Table \ref{tab:lit_z9_ew}). The PRIMAL
$z=5-7$, and $z=7-9$ subsamples have median values of $\log({\rm EW}_{\rm
rest}^{\rm H\beta+[O\textsc{iii}]}/\si{\angstrom}) = 3.11\pm 0.29$ and
$\log({\rm EW}_{\rm rest}^{\rm H\beta+[O\textsc{iii}]}/\si{\angstrom}) =
3.06\pm 0.19$, respectively, which are consistent with the median value
inferred for the combined $z\geq 9$ sample. Moreover, from a 2-sided
Kolmogorov-Smirnov test, we find that the two PRIMAL EW-distributions are
indistinguishable from the $z\geq 9$ distribution.

\smallskip

The observed EW distribution of our combined $z\geq 9$ sample and its median
value are based on direct EW measurements, and do not take into account the EW
uncertainties for each galaxy. In order to fold in these uncertainties and to
try and characterize the underlying distribution of H$\beta$+[O\textsc{iii}]
rest-frame EWs that gives rise to our observed EW values, we follow the method
described in \cite{Endsley2021} (see also \cite{Boyett2022}). The underlying
distribution of EW-values is assumed to be log-normal, characterized by the
median, $\mu_{\rm EW}$, and the standard deviation, $\sigma_{\rm EW}$, and the
goal is to find the most likely values of these two parameters given our data.
Following \cite{Endsley2021}, we construct a grid covering $\log_{10}(\mu_{\rm
EW}/\si{\angstrom})$ ranging from $1.0$ to $3.5\,{\rm dex}$ and $\sigma_{\rm
EW}$ varying between $0.01$ and $1.0\,{\rm dex}$, utilizing a uniform spacing
of $0.01\,{\rm dex}$ for both parameters. Subsequently, at each grid-point we
calculate the probability, $P(\mu_{\rm EW}, \sigma_{\rm EW})$ for that set of
parameters: $P(\mu_{\rm EW}, \sigma_{\rm EW}) \propto \prod_{i} P_i({\rm EW})
P({\rm EW} | \mu_{\rm EW}, \sigma_{\rm EW})$. Here, $P_i({\rm EW})$ is the
probability distribution function of the EW-value measured for the $i$th
source. For our MIDIS sample, this is given by the EW-distributions used to
derive the EW-errors as described in \S\ref{subsection:EW-values}. For the
literature sources, we ignore the two sources with only upper limits on ${\rm
EW}_{\rm rest}^{\rm H\beta+[O\textsc{iii}]}$, and for the remaining 14 sources
we adopt a normal distribution centered on the measured EW-values and with a
standard deviation corresponding to the reported errors. We derive a median EW
of $\log(\mu_{\rm EW}/\si{\angstrom}) = 3.19^{+0.09}_{-0.12}$ (i.e., $\mu_{\rm
EW} = 1550^{+357}_{-374}\,\si{\angstrom}$) and a standard deviation of
$\sigma_{\rm EW} = 0.31^{+0.09}_{-0.07}\,{\rm dex}$. 

Finally, we also compare with the EW distribution of simulated $z= 9-11$
galaxies selected from the First Light and Reionisation Epoch Simulations
\citep[{\tt FLARES};][]{Lovell2021,Vijayan2021}. Along with various galaxy
properties, the {\tt FLARES} simulations predict line luminosities and
equivalent widths for the most prominent nebular lines, including H$\beta$ and
[O\textsc{iii}], as well as fluxes in all the {\it HST} and {\it JWST}
broadband filters. In {\tt FLARES} the emission lines are implemented using the
{\tt CLOUDY} code \citep{Ferland2017}, assuming an ionisation parameter, $U$,
that is scaled with a reference value ($U_{\rm ref} =0.01$). Spherical,
ionisation-bound nebulae are assumed with a gas density of $10^{2.5}\,{\rm
cm^{-3}}$. Importantly, the EW-values provided by {\tt FLARES} are calculated
relative to the total, i.e., stellar {\it and} nebular, continuum. In order to
facilitate a fair comparison between the {\tt FLARES} and the observed $z\geq
9$ (MIDIS+literature) EW distributions, we first restrict the simulated sample
to galaxies in the redshift range $z=9-11$ and, secondly, they must be
detectable by the MIDIS survey, i.e., we only include sources with F560W
magnitudes brighter than $28.59$. Also, we require their
H$\beta$+[O\textsc{iii}] EWs to be $\ge 60\,\si{\angstrom}$ in order to match
the minimum EW corresponding to the adopted flux excess critation of $\Delta m
\leq -0.2$ (see \S\ref{subsection:EW-values}). This resulted in 3258 galaxies
with an EW distribution as the black curve in Fig.\,\ref{fig:ew_hist}. The {\tt
FLARES} EW distribution has a median value of $\log({\rm EW}^{\rm
H\beta+[O\textsc{iii}]}_{\rm rest}/\si{\angstrom}) = 2.94\pm 0.11$. The
observed median is therefore higher by $0.16\,{\rm dex}$ (a factor of $\sim
1.5$). Although the distributions partially overlap, this offset in the medians
suggest they are not statistically consistent. A bootstrap test drawing mock
samples of equal size from the {\tt FLARES} EW distribution shows that
obtaining a median as high as observed occurs with probability $p = 5\times
10^{-6}$, indicating that the observed $z\geq 9$ sample (MIDIS+literature) is
statistically inconsistent with being drawn from the {\tt FLARES} parent
population. We note, however, that differences in sample selection, the
specific method adopted for EW measurements, and the treatment of extreme
emission-line systems in {\tt FLARES} may contribute to the observed offset.

While it is instructive to examine the EW distributions of strong
H$\beta$+[O\textsc{iii}] line emitters identified in slightly different ways
from various surveys, any comparison should be cautioned by the fact that the
surveys probe a wide range of redshifts, UV luminisities and stellar masses.
Also, several studies have now shown the existence of non-negligible trends
between  H$\beta$+[O\textsc{iii}] EW and UV luminosity and stellar mass
\citep[e.g.,][]{Endsley2021, Endsley2024}. In the following sections, we will
examine these trends in the context of our sample.

\subsection{H$\beta$+[O\textsc{iii}] rest-frame EW vs UV luminosity}
Samples of strong H$\beta$+[O\textsc{iii}] emitters identified via flux-excess
techniques have now been studied out to $z \simeq 9$, and several recent
analyses report a dependence of the EW distribution on UV luminosity.
\citet{Endsley2024} analysed 759 galaxies at $z \simeq 6-9$ in JADES and found
that both $\mu_{\rm EW}$ and $\sigma_{\rm EW}$ depend systematically on $M_{\rm
UV}$. In their sample, brighter galaxies exhibit higher median equivalent
widths, while the dispersion increases toward fainter magnitudes. At $z \simeq
6$, their ``bright'' ($M_{\rm UV} \simeq -20$), ``faint'' ($M_{\rm UV} \simeq
-18.7$), and ``very faint'' ($M_{\rm UV} \simeq -17.5$) subsamples have median
EWs of $890$, $590$, and $380\,{\rm \AA}$, respectively, with corresponding
dispersions of $0.31$, $0.37$, and $0.51\,{\rm dex}$. A similar trend is
observed at $z \simeq 7-9$, where the median EW declines and the width of the
distribution increases toward lower luminosities. This behaviour, i.e., a
higher typical EW but reduced scatter in UV-bright systems, is interpreted as
being as consistent with combination of lower metallicity and increasingly
bursty star-formation histories in UV-faint galaxies (and thus a more even
ratio of SFR-rising vs SFR-declining galaxies, compared to the UV-bright
galaxies). \citet{Begley2025} reports consistent behaviour in $\mu_{\rm EW}$
and $\sigma_{\rm EW}$ with $M_{\rm UV}$ for 279 PRIMER+JADES galaxies at $z
\simeq 6.9-7.6$. Dividing their sample into three equally sized $M_{\rm UV}$
bins ($\langle M_{\rm UV}\rangle \simeq -19.9$, $-19.3$, and $-18.3$), they
measure a $\simeq 0.17\,{\rm dex}$ increase in $\mu_{\rm EW}$ from $M_{\rm UV}
\simeq -18$ to $-20$, corresponding to $d{\rm EW}/d M_{\rm UV} \sim -140\,{\rm
\si{\angstrom}\,mag^{-1}}$. They also find that the width of the EW
distribution decreases toward brighter galaxies. 

\begin{figure}
    \centering
    \includegraphics[width=\linewidth]{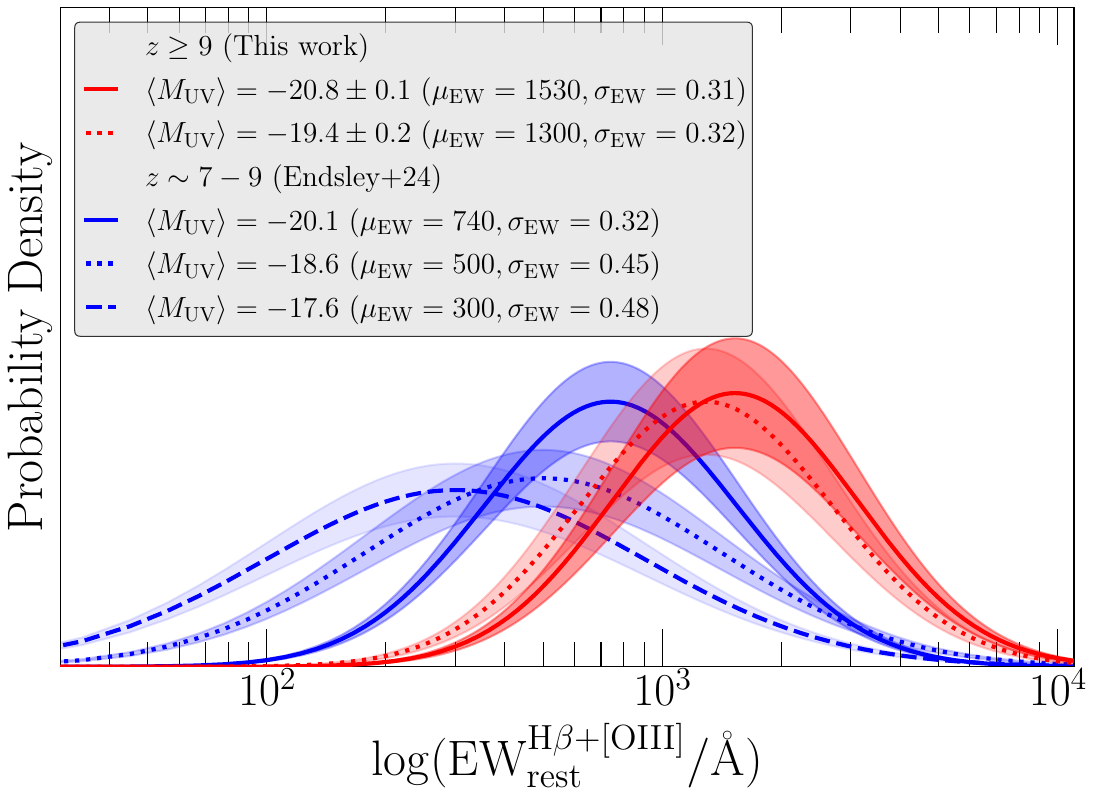}
    \caption{
    Inferred H$\beta$+[O\textsc{iii}] EW distribution for our the bright and
    faint subsets of our combined $z\geq 9$ (MIDIS + literature) sample (red
    solid amd dashed distributions, respectively), as described in
    \S\ref{subsection:EW-values}. The blue solid, dotted and dashed curves show
    similarly derived H$\beta$+[O\textsc{iii}] EW distributionsn for $z\sim
    7-9$ galaxies with $\langle M_{\rm UV}\rangle = -20.1, -18.6$, and $-17.6$,
    respectively, from the JADES survey \citep{Endsley2024}.
    }
    \label{fig:ew_pdf}
\end{figure}

Using the luminosity-binned medians reported by \citet{Endsley2024}, we can
derive $d {\rm EW}/d M_{\rm UV}$ in a similar manner as \citet{Begley2025}.
Their $z \simeq 6$ sample implies an evolution of $\simeq 0.30\,{\rm dex}$ in
$\mu_{\rm EW}$ between $M_{\rm UV} \simeq -18$ and $-20$, corresponding to
$d{\rm EW}/d M_{\rm UV} \simeq -220\,{\rm \si{\angstrom}\,mag^{-1}}$. For their
$z \simeq 7-9$ sample, we find a similar scaling, $\simeq 0.31\,{\rm dex}$ over
the same range,  corresponding to $d {\rm EW}/d M_{\rm UV} \simeq -195\,{\rm
\si{\angstrom}\,mag^{-1}}$. These slopes are somewhat steeper than the
evolution ($d {\rm EW}/d M_{\rm UV} \sim -140\,{\rm \si{\angstrom}\,mag^{-1}}$)
reported by \citet{Begley2025}, but all consistently indicate stronger nebular
emission in brighter galaxies at $z \simeq 6-8$.

Our full $z \geq 9$ sample spans $M_{\rm UV} = -21.8$ to $-18.8$, with a median
of $-20.45 \pm 0.56$ (median absolute deviation, MAD). We split our sample
into: {\it i)} a bright ($M_{\rm UV} \leq -20.5$) sub-sample consisting of 9
sources and have $\langle M_{\rm UV}\rangle = -20.8\pm 0.1$, and {\it ii)} a
faint ($M_{\rm UV} \geq -20.5$) sub-sample, consisting of  7 sources (including
our three MIDIS sources) and have $\langle M_{\rm UV}\rangle = -19.4\pm 0.2$.
Modeling the subsamples separately under the assumption of log-normal EW
distributions (as described in \S\ref{subsection:first-look}) yields $\mu_{\rm
EW} = 1530^{+180}_{-49}\,{\rm \si\angstrom}$ for the bright subsample, with
$\sigma_{\rm EW} = 0.31^{+0.01}_{-0.01}\,{\rm dex}$, and $\mu_{\rm EW} =
1300^{+210}_{-55}\,{\rm \si{\angstrom}}$ for the faint subsample, with
$\sigma_{\rm EW} = 0.32^{+0.01}_{-0.01}\,{\rm dex}$. Repeating this
inference-analysis of $\mu_{\rm EW}$ and $\sigma_{\rm EW}$ multiple times in
order to assess the robustness of our results, we find the bright subsample
consistently yields slightly larger $\mu_{\rm EW}$-values than the faint
sub-sample. The offset is $\Delta \mu_{\rm EW} \simeq 230\,\si{\angstrom}$
($0.06\,{\rm dex}$), corresponding to a $\sim 18$ per cent increase in the
median EW. In contrast, the inferred $\sigma_{\rm EW}$-values are similar in
the two subsamples. This $230\,\si{\angstrom}$ increase in $\mu_{\rm EW}$ over
$\Delta M_{\rm UV} \simeq 1.4\,{\rm mag}$ corresponds to $d {\rm EW}/d M_{\rm
UV} \simeq -164\,{\rm \si{\angstrom}\,mag^{-1}}$). This slope is consistent
with the trends inferred at $z \simeq 6-9$ \citep{Endsley2024, Begley2025}.
However, we we stress that our derived slope is sensitive to small-number
statistics and the limited coverage in $M_{\rm UV}$. Moreover, we note that the
$\mu_{\rm EW}$ values derived from our bright and faint $z\geq 9$ subsamples
are significantly higher than the values derived for the $M_{\rm
UV}$-corresponding subsamples from \citet{Endsley2024}. We attribute this to
selection effects and the incompleteness of the $z\geq 9$ sample.

As already mentioned, we do not find any evidence of an increasing $\sigma_{\rm
EW}$ toward lower luminosities, as reported by \citet{Endsley2024} and
\citet{Begley2025}. Our bright and faint $z \geq 9$ subsamples show similar
scatter ($\sigma_{\rm EW} \simeq 0.30-0.32\,{\rm dex}$), which are similar to
the scatter found in UV-bright galaxies at $z\simeq 7-9$ \citep{Endsley2024}.
This may indicate that the luminosity-dependent broadening of the EW
distribution seen at $z \simeq 6-9$ is not yet firmly established at $z \geq
9$, and that star-formation is equally stochastic, and significant, across the
general $\geq 9$ galaxy population. Alternatively, the limited sample size and
limited $M_{\rm UV}$ distribution may obscure an underlying shallow trend. 

Thus, while a coherent ${\rm EW}_{\rm rest}^{\rm H\beta+[O\textsc{iii}]}-M_{\rm
UV}$ relation in both the median and dispersion appears to be established at $z
\simeq 6-9$ \citep{Endsley2024, Begley2025},  our $z \geq 9$ sample does not
reveal a statistically unambiguous dependence in either quantity. Larger, more
uniformly sampled datasets will be required to determine whether the
luminosity-dependent evolution in both the typical EW and its intrinsic scatter
persists into the earliest stages of galaxy assembly.

\bigskip

\subsection{H$\beta$+[O\textsc{iii}] rest-frame EW vs stellar mass}
\begin{figure*}
    \centering
    \includegraphics[width=1.0\linewidth]{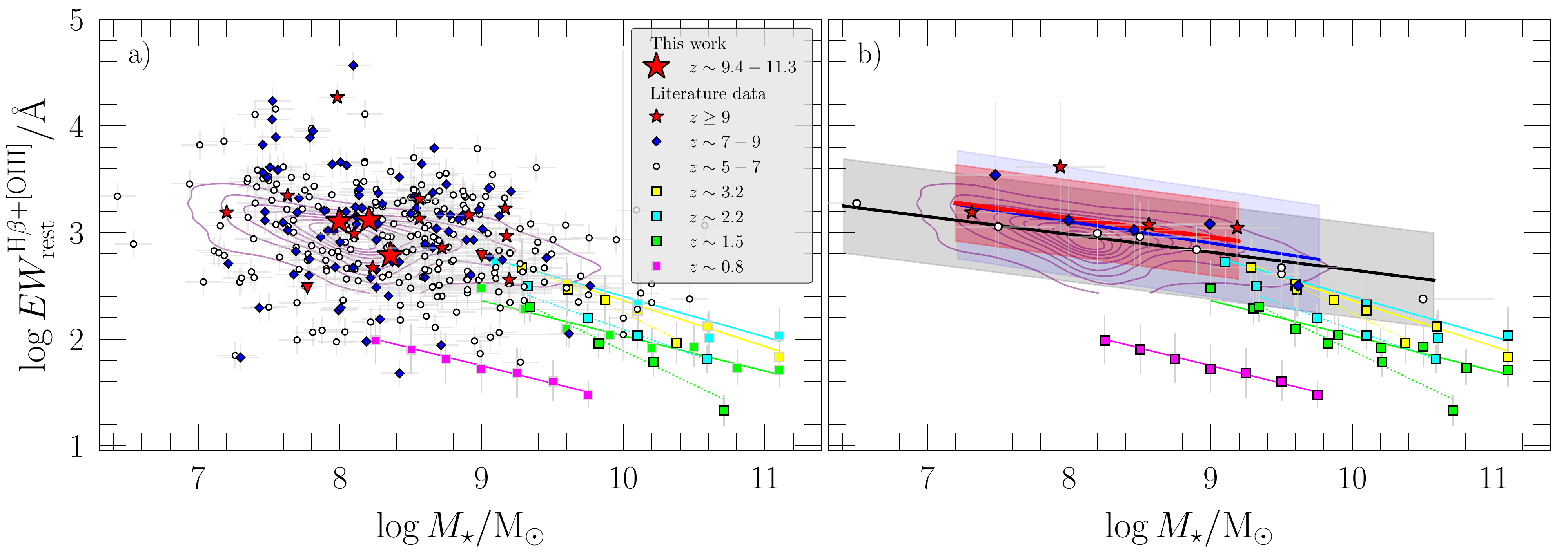}
    \caption{
        {\bf a)} H$\beta$+[O\textsc{iii}] rest-frame EW as a function of
        stellar mass for our $z\simeq 9.4-11.3$ MIDIS sample (large red stars)
        and $z\geq 9$ sources from the literature (small red stars; see Table
        \ref{tab:lit_z9_ew}). For comparison, we include samples at $z=7-9$
        \citep{Endsley2021, Rinaldi2023, Heintz25} and $z=5-7$
        \citep{Endsley2021, Endsley2023b, Rinaldi2023, Heintz25}. At $z\ls
        3.2$, we show binned measurements from the literature
        \citep{Schenker2013, Khostovan2016, Reddy2018} together with their
        published log-linear fits (solid lines). Purple contours indicate the
        density distribution of $z\simeq 9-11$ galaxies from the {\tt FLARES}
        simulations \citep{Lovell2021,Vijayan2024} that satisfy the same
        selection criteria as the MIDIS sample. {\bf b)} Same as panel (a), but
        with the $z=5-7$, $z=7-9$, and $z\geq 9$ samples binned in stellar mass.
        Log-linear fits to the individual data-points (i.e., non-binnned) data are
        shown as solid lines, with the corresponding residual scatter indicated by the
        shaded regions.
        }
    \label{fig:EW-vs-mstar}
\end{figure*}

In Fig.\,\ref{fig:EW-vs-mstar}a, we plot the H$\beta$+[O\textsc{iii}]
rest-frame EW versus stellar mass for our $z\sim 9.4-11.3$ MIDIS sample (large
red stars) alongside literature measurements spanning $z \simeq 0.8-11$
\citep{Khostovan2016, Reddy2018, Endsley2021, Endsley2023a, Endsley2023b,
Rinaldi2023, Heintz25}. In addition, we include the $z\ge 9$ compilation (small
red stars or red triangles for upper limits) assembled in
\S\ref{section:literature-sample} (Table \ref{tab:lit_z9_ew}), which provides
the most direct comparison sample to MIDIS. Literature data are grouped into
redshift bins and colour-coded: the $z \simeq 0.8$ bin is from the HiZEL survey
\citep{Khostovan2016}, while the $z \simeq 1.5-3.2$ bins combine results from
the HiZEL \citet{Khostovan2016} and MOSDEF \citet{Reddy2018} surveys. Square
symbols and error bars mark the median and r.m.s. scatter in each stellar mass
bin; solid and dotted lines show the respective power-law fits from the two
studies. Although both cover similar stellar mass and redshift ranges,
\citet{Reddy2018} tend to report lower EWs, particularly at $z \simeq 2.2$.
Above $z \simeq 5$, the symbols in Fig. \ref{fig:EW-vs-mstar}a correspond to
individual galaxies, colour-coded according to the redshift bins $z=5-7$,
$7-9$, and $\geq 9$. The $z=5-9$ EW data are derived from both broadband excess
(\citealt{Endsley2021, Endsley2023a, Endsley2023b, Rinaldi2023}) as well as
spectroscopy \citep{Heintz25}, as are the $z\ge 9$ points (although primarily
spectroscopy, see \S\ref{section:literature-sample}).

At $z \ls 4$, several studies have found an inverse correlation between EW and
stellar mass \citep[e.g.,][]{Khostovan2016, Reddy2018}. This is expected if EW
scales inversely with continuum flux, which increases with stellar mass.
Differences in slope with redshift suggest additional influences from dust
attenuation and metallicity. Inverse ${\rm EW} - M_{\star}$ relationships have
been observed for other optical emission lines, e.g., H$\alpha$ and
[O\textsc{ii}], although the strongest correlation is seen for
H$\beta$+[O\textsc{iii}] \cite[e.g.,][]{Reddy2018}. For $z \ls 4$, the slope
remains roughly constant while the normalisation increases with redshift; for
instance, at $M_\star = 10^{9.7} M_\odot$, the median EW at $z \simeq 2.3$ is
about $30\times$ that at $z \simeq 0$ \citep{Reddy2018}. This evolution is
visible in Fig.\,ref{fig:EW-vs-mstar}, where we plot the average ${\rm EW}_{\rm
rest}^{\rm H\beta+[O\textsc{iii}]}-M_{\star}$ relations from the HiZEL
\citep{Khostovan2016} and MOSDEF \citep{Reddy2018} surveys at $z \simeq 0.8,
1.5, 2.2, 3.2$.  

Fig.\,\ref{fig:EW-vs-mstar}a shows that this inverse ${\rm EW}_{\rm rest}^{\rm
H\beta+[O\textsc{iii}]} - M_{\star}$ relation extends to $z \simeq 5-7$, and $z
\simeq 7-9$. Applying both a Pearson linear correlation test and a Spearman
rank test to the 268 galaxies in the $z=5-7$ bin yields highly significant
anti-correlation coefficients $r_{\rm S}\sim -0.33$ and $p$-values $< 10^{-8}$.
The $z = 7-9$ sample, which consists of 109 galaxies, shows only marginal
evidence of a ${\rm EW}_{\rm rest}^{\rm H\beta+[O\textsc{iii}]} - M_{\star}$
anti-correlation. The Pearson test gives $r = -0.21$ ($p = 0.044$), while the
rank-based Spearman test suggest that the correlation is at best marginal
($r_{\rm S} = -0.18$, $p = 0.086$). Log-linear fits to the individual galaxies
in the two samples yield very similar results: $\log({\rm EW}_{\rm rest}^{\rm
H\beta+[O\textsc{iii}]}/\si{\angstrom}) = (-0.20\pm
0.01)\log(M_{\star}/\Msolar) + (4.59\pm 0.09)$ (with an r.m.s. scatter about
the fit of $\sigma_{\rm res}=0.44\,{\rm dex}$) for the $z=5-7$  bin and
$\log({\rm EW}_{\rm rest}^{\rm H\beta+[O\textsc{iii}]}/\si{\angstrom}) =
(-0.18\pm 0.02)\log(M_{\star}/\Msolar) + (4.53\pm 0.19)$ (with an r.m.s.
scatter about the fit of $\sigma_{\rm res}=0.51\,{\rm dex}$) for the $z=7-9$
bin (shown as black and blue lines, respectively, in
Fig.\,\ref{fig:EW-vs-mstar}b). Recent {\it JWST}-based studies reach similar
conclusions at $z \gs 5$ using different selections and methodologies
\citep[e.g.,][]{Matthee2023, Rinaldi2023, Caputi2024, Begley2025}.
\citet{Matthee2023} show that spectroscopically confirmed {\rm [O\textsc{iii}]}
emitters at $z\simeq 5.3-7.0$ display much larger {\rm [O\textsc{iii}]} EWs at
low stellar mass, effectively extending the ${\rm EW}_{\rm rest}^{\rm
H\beta+[O\textsc{iii}]}-M_{\star}$ trend into the reionization era. Similarly,
\citet{Caputi2024} found a broad inverse ${\rm EW}_{\rm rest}^{\rm
H\beta+[O\textsc{iii}]}-M_{\star}$ for $z\simeq 5.5-8$ galaxies. They argued
that part of the trend is due to correlation between stellar mass and
population age, while simultaneously emphasizing that a substantial fraction of
photometrically selected galaxies at these redshifts exhibit weak
H$\beta$+[O\textsc{iii}] emission (i.e., ${\rm EW}_{\rm rest}^{\rm H\beta +
[O\textsc{iii}]} < 100\,\si{\angstrom}$) and would be missed by line-excess
selections. This emphasizes that selection and completeness corrections can
modify the apparent slope at the lowest masses, since low-mass galaxies are
preferentially identified when they host stronger emission lines \citep[see
also][]{Begley2025}.

At $z\gs 9$, the combined MIDIS and $z\geq 9$ literature compilation shows that
H$\beta$+[O\textsc{iii}] emission persists across a fairly broad range in
stellar mass. This sample has negative Spearman and Kendall rank coefficients,
indicating a negative trend is present between $\log {\rm EW}$ and $\log
M_\star$. The correlation is not statistically significant, however. A
log-linear fit to the combined sample (shown as the red line in
Fig.\,\ref{fig:EW-vs-mstar}b) yields $\log({\rm EW}_{\rm rest}^{\rm
H\beta+[O\textsc{iii}]}/\si{\angstrom}) = (-0.17\pm
0.04)\times\log(M_{\star}/\Msolar) + (4.56\pm 0.30)$  (with an r.m.s. scatter
about the fit of $\sigma_{\rm res}=0.36\,{\rm dex}$), in good agreement with
the $z=5-7$ and $7-9$ samples.

The distribution of simulated $z=9-11$ {\tt FLARES} galaxies in the ${\rm
EW}_{\rm rest}^{\rm H\beta+[O\textsc{iii}]} - M_{\star}$ plane is shown as
purple contours in Fig.\,\ref{fig:EW-vs-mstar}a and b. The simulated galaxies
exhibit an anti-correlation between ${\rm EW}_{\rm rest}^{\rm
H\beta+[O\textsc{iii}]}$ and $M_{\star}$. A log-linear fit to the simulated
galaxies yields: $\log({\rm EW}_{\rm rest}^{\rm
H\beta+[O\textsc{iii}]}/\si{\angstrom}) = (-0.18\pm
0.03)\times\log(M_{\star}/\Msolar) + (4.41\pm 0.25)$, which is consistent with
the relations fitted to the $z = 5-7$ and $7-9$ samples. While the simulated
galaxies span an EW-range of approximately $50-3000\,\si{\angstrom}$, and
overlap significantly in the parameter space with the MIDIS and $z\geq 9$
literature sources, a slight offset from the observations is discernible. This
is expected given that in \S\ref{subsection:EW-hist} we demonstrated that the
EW distribution of $z=9-11$ {\tt FLARES} galaxies peak at somewhat lower
values, compared to our combined $z\geq 9$ sample. Moreover, the simulations do
not reproduce the most extreme EW-values ($\log({\rm EW}_{\rm rest}^{\rm
H\beta+[O\textsc{iii}]}/\si{\angstrom}) \gs 3.4$) observed, with the maximum
simulated value reaching $2975\,\si{\angstrom}$. 

\subsection{Redshift evolution of the H$\beta$+[O\textsc{iii}] rest-frame EW}
\begin{figure*}
    \centering
    \includegraphics[width=0.9\linewidth]{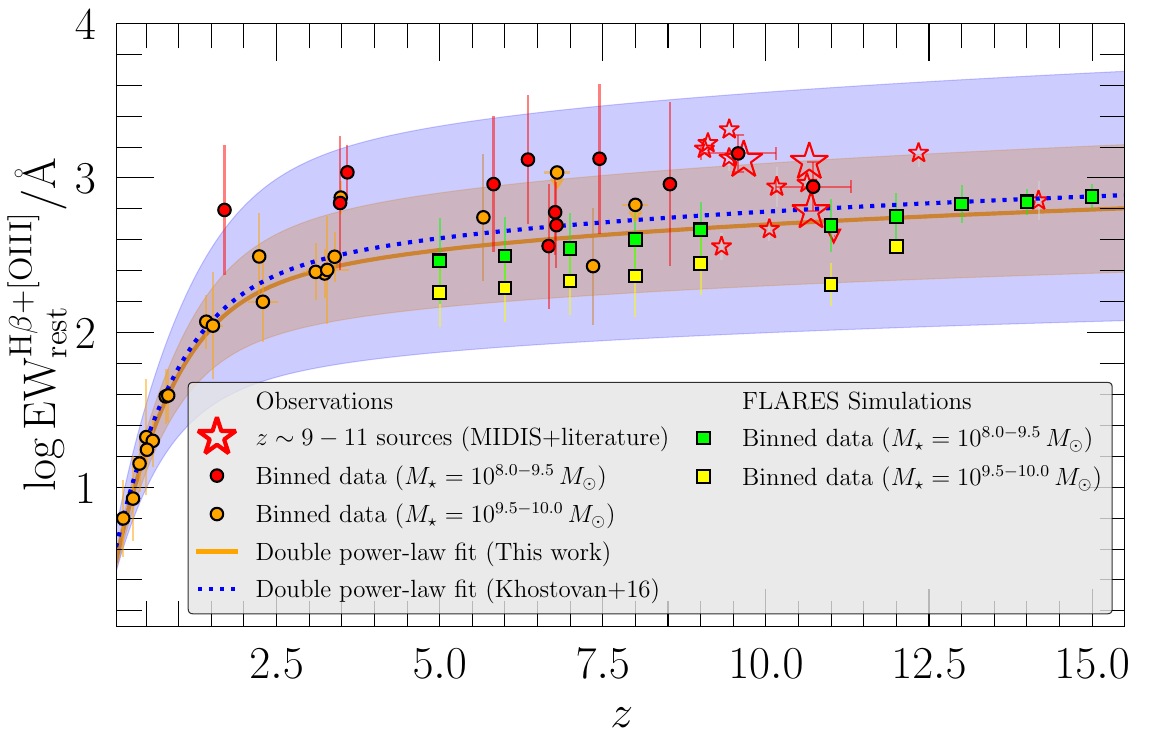}
    \caption{
        H$\beta$+[O\textsc{iii}] restframe EW measurements of star-forming
        galaxy samples as a function of redshift. The large open red stars
        highlight our $z\simeq 9.4-11.3$ MIDIS sample listed in Table
        \ref{tab:candidate-list} and the small open red stars the $z\geq 9$
        literature sample (Table \ref{tab:lit_z9_ew}. The red circles show
        average H$\beta$+[O\textsc{iii}] restframe EW measurements of galaxy
        samples with stellar masses in the range $\log(M_{\star}/\Msolar) =
        8.0-9.5$, and are compiled from spectroscopic as well as photometric
        surveys in the literature
        \citep{Schenker2013,Holden2016,Endsley2021,Rinaldi2023, Heintz25} The
        orange circles indicate galaxy samples with stellar masses in the range
        $\log(M_{\star}/\Msolar) = 9.5-10.0$, also compiled from the literature
        \citep{Lamareille2009,Thomas2013,Labbe2013,Schenker2013,Smit2015,Khostovan2016,Holden2016,Endsley2021}.
        The orange curve and shaded region show a double power-law fit,and the
        associated 95\% confidence intervals, to these data from the literature.
        }
    \label{fig:EW-vs-z}
\end{figure*}

Fig.\,\ref{fig:EW-vs-z} shows the evolution of the H$\beta$+[O\textsc{iii}]
rest-frame EW as a function of redshift, combining our new $z\simeq 9.4-11.3$
MIDIS measurements, our compiled $z\geq 9$ sample (Table \ref{tab:lit_z9_ew}),
and literature data spanning $z\simeq 0-8$ \citep{Lamareille2009, Thomas2013,
Labbe2013, Schenker2013, Stark2014, Smit2015, Holden2016, Khostovan2016,
Malkan2017, Reddy2018, Endsley2021, Endsley2023a, Rinaldi2023, Heintz25}. The
circles in Fig.\,\ref{fig:EW-vs-z} represent sample-averages in bins of
redshift, and are further divided into a high-mass sample
($\log(M_{\star}/\Msolar) = 9.5-10.0$; orange circles) and a low-mass sample
($\log(M_{\star}/\Msolar) = 8.0-9.5$; red circles). The latter matches the
stellar mass-range of our sample with the exception of source ID 2987 (see
Table \ref{tab:candidate-list}).

Following \cite{Khostovan2016}, we fit a double power-law of the form ${\rm
EW}(z) = {\rm EW}(z=0) (1+z)^{\gamma} / [1 + [(1+z)/c]^{\epsilon}]$ to the
high-mass sample. We find ${\rm EW}(z=0) = 2.85\pm 0.33$, $\gamma = 4.73\pm
0.33$, $c = 2.48\pm 0.22$, and $\epsilon = 4.14\pm 0.23$. The resulting curve
(orange line in Fig.\,\ref{fig:EW-vs-z}) and its 95\% confidence intervals are
fully consistent with the original \cite{Khostovan2016} fit (blue curve),
showing a rapid rise from $z\sim 0$ to $z\sim 2-3$, followed by a flattening at
$z\gs 3$, in agreement with \cite{Reddy2018}. For the low-mass sample, all
available H$\beta$+[O\textsc{iii}] measurements lie at $z\gs 1.5$, and the
uncertainties are larger owing to smaller sample sizes. We therefore do not
attempt a double power-law fit. Nevertheless, between $z\sim 1.5-3$, the
low-mass galaxies systematically exhibit higher average
H$\beta$+[O\textsc{iii}] EWs than the high-mass galaxies, as expected from the
inverse ${\rm EW}-M_{\star}$ relation discussed in the previous section.

Our MIDIS sample, combined with the compiled $z\geq 9$ literature sample
extends measurements of H$\beta$+[O\textsc{iii}] rest-frame EWs into this still
poorly explored redshift regime. The combined $z\geq 9$ dataset occupies the
same region of ${\rm EW}-z$ parameter space as the low-mass galaxies at $z\sim
5-8$. Importantly, we find no evidence for a renewed steep rise in EW beyond
$z\sim 9$. Instead, the typical EW values at $z\geq 9$ remain broadly
consistent with the plateau established at $z\gs 3$ \citep{Khostovan2016,
Reddy2018}, falling within the envelope defined by the extrapolated double
power-law fits. Within current uncertainties, we therefore find no
statistically significant indication of either a dramatic upturn in EW at $z\gs
9$, nor a systematic decline relative to the $z\sim 5-8$ populations. This
result is particularly noteworthy given theoretical expectations that
decreasing metallicity, evolving ionization conditions, or extremely young
stellar populations at these early epochs could substantially modify rest-frame
optical line strengths \citet[e.g.,][]{Trussler2024}. Instead, taken at face
value, our analysis suggests that the physical processes regulating nebular
emission at  $z\gs 9$ may represent a continuation of trends already
established by $z\sim 5-8$. To fully determine whether subtle evolution in the
normalization or scatter of the ${\rm EW}-z$ relation emerges at $z \gs 9$
would require improved statistics and more uniformly selected samples.

The {\tt FLARES} simulations span a redshift range from $z=15$ to $z=5$, and in
Fig.\,\ref{fig:EW-vs-z} we show the median EW-values at redshifts $z=5, 6, ...,
15$ (in steps of $\Delta z = 1$) for galaxies falling in the mass-bins
$\log(M_{\star}/\Msolar) = 8.0 - 9.5$ (green squares) and
$\log(M_{\star}/\Msolar) = 9.5 - 10.0$ (yellow squares). The simulations
reproduce the overall increase of EW with redshift, although they predict
systematically lower normalisations than observed. As in the observations, the
simulated galaxies with lower stellar masses have higher median EWs than the
more massive galaxies, consistent with the inverse ${\rm EW}-M_{\star}$
relation inferred from observations. It is important to note that, in this
comparison, we have not imposed any observationally motivated selection on the
{\tt FLARES} galaxies; instead, we use the full simulated population. This
difference likely explains the lower normalisation of the simulated ${\rm
EW}-z$ relation, since both observed flux-excess and spectroscopic selection
preferentially targets systems with stronger emission lines. Consequently, {\tt
FLARES} likely represent the intrinsic galaxy population, whereas the observed
samples are biased toward higher EWs.

\subsection{Constraints on the $z\sim 9.4-11.3$ H$\beta$+[O\textsc{iii}] luminosity function}\label{section:LF}
An increasing number of studies, although still relatively few, have made
estimates of the H$\beta$+[O\textsc{iii}] luminosity function up to $z\sim 5-8$
\citep{DeBarros2019,Matthee2023,Meyer2024,Wold2025,Korber2025,Meyer2025}.
Beyond $z\sim 9$, however, no attempts have been made. Here, we will use our
sample to put the first  direct constraints on the  H$\beta$+[O\textsc{iii}]
luminosity function at $z > 9$, with the obvious caveats of small number
statistics and cosmic variance. 

In order to estimate the H$\beta$+[O\textsc{iii}] luminosity function across
the redshift range $z\sim 9.4-11.3$, we use the H$\beta$+[O\textsc{iii}] line
luminosities and the associated uncertainties of our sample galaxies (see Table
\ref{tab:candidate-list}). The luminosity function was derived by adopting a
non-parametric $1/V_{\rm max}$ method \citep{Efstathiou1988}. All the sources
in our sample have F560W AB magnitudes brighther than the 5-$\sigma$ depth of
the shallowest part of the MIDIS F560W image ($m_{\rm 5.6\mu m} \sim 27.68$,
see \citet{Oestlin2025}). We therefore expect the completeness of our sample to
be at least 85-90\%, which is also confirmed by an extensive completeness
analysis of the MIDIS field \citep{Jermann2026}. Owing to the small survey area
of MIDIS, cosmic variance must be included in the uncertainty budget. We
estimate this term by scaling from the empirically calibrated cosmic variance
measured in the COSMOS-3D survey \citep{Meyer2025}, which covers $0.3\,{\rm
deg}^2$ at $z \simeq 7-9$ and finds a fractional variance of $\sigma_{\rm cv}
\simeq 0.15$ in the number density of H$\beta$+[O\,\textsc{iii}] emitters.
Assuming that the variance scales inversely with the square root of the
surveyed area for a fixed tracer population, we obtain $\sigma_{\rm cv} = 0.15
\sqrt{\frac{1080}{4.7}} \simeq 2.27$, corresponding to a $\sim 230\%$
fractional uncertainty for our field. We incorporate this by adding the
cosmic-variance term in quadrature to the Poisson uncertainty in each
luminosity bin, $\left(\frac{\delta\Phi}{\Phi}\right)^2 =
\left(\frac{\delta\Phi}{\Phi}\right)_{\rm Poisson}^2 + \sigma_{\rm cv}^2$. The
resulting estimate of the H$\beta$+[O\textsc{iii}] luminosity function across
the redshift range $z\sim 9.4-11.3$ is $\Phi \sim 10^{-3.4}\,{\rm
Mpc^{-3}\,dex^{-1}}$ at $\log(L_{\rm H\beta+[O\textsc{iii}]}/{\rm
erg\,s^{-1}})\sim 42.5$. 
\begin{figure}
    \centering
    \includegraphics[width=1.0\linewidth]{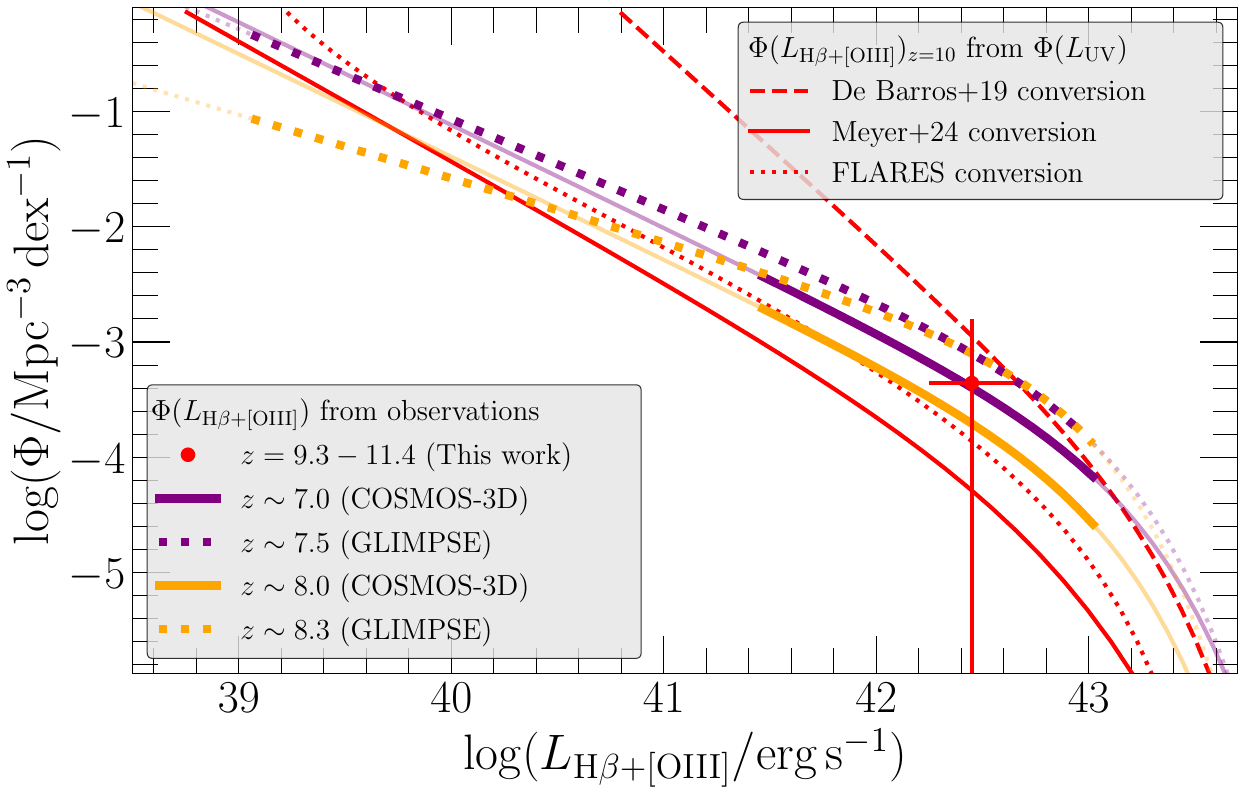}
    \caption{
        H$\beta$+[O\textsc{iii}] luminosity function estimates at $z\simeq
        9.4-11.3$ (red circles, this work). Also shown are determinations of
        the H$\beta$+[O\textsc{iii}] luminosity functions at $z\sim 7.0-7.5$
        and $\sim 8.0-8.3$, shown as purple and yellow curves, respectively,
        based on unbiased spectroscopic surveys from \citet{Meyer2025} (solid
        curves) and \citet{Korber2025} (dotted curves). The thick parts of the curves
        indicate the luminosity range directly probed by the surveys. The red solid,
        dashed and dotted curves show the derived $z\sim 10$ H$\beta$+[O\textsc{iii}]
        luminosity functions based on the UV luminosity function at $z\sim 10$ from
        \citet{Whitler2025} and three empirical $L_{\rm H\beta+[O\textsc{iii}]} -
        L_{\rm UV}$ conversion (see \S\ref{section:LF} and
        Fig.\,\ref{fig:LOIIIHb_LUV}).
        }
    \label{fig:LF}
\end{figure}
\begin{figure}
    \centering
    \includegraphics[width=1.0\linewidth]{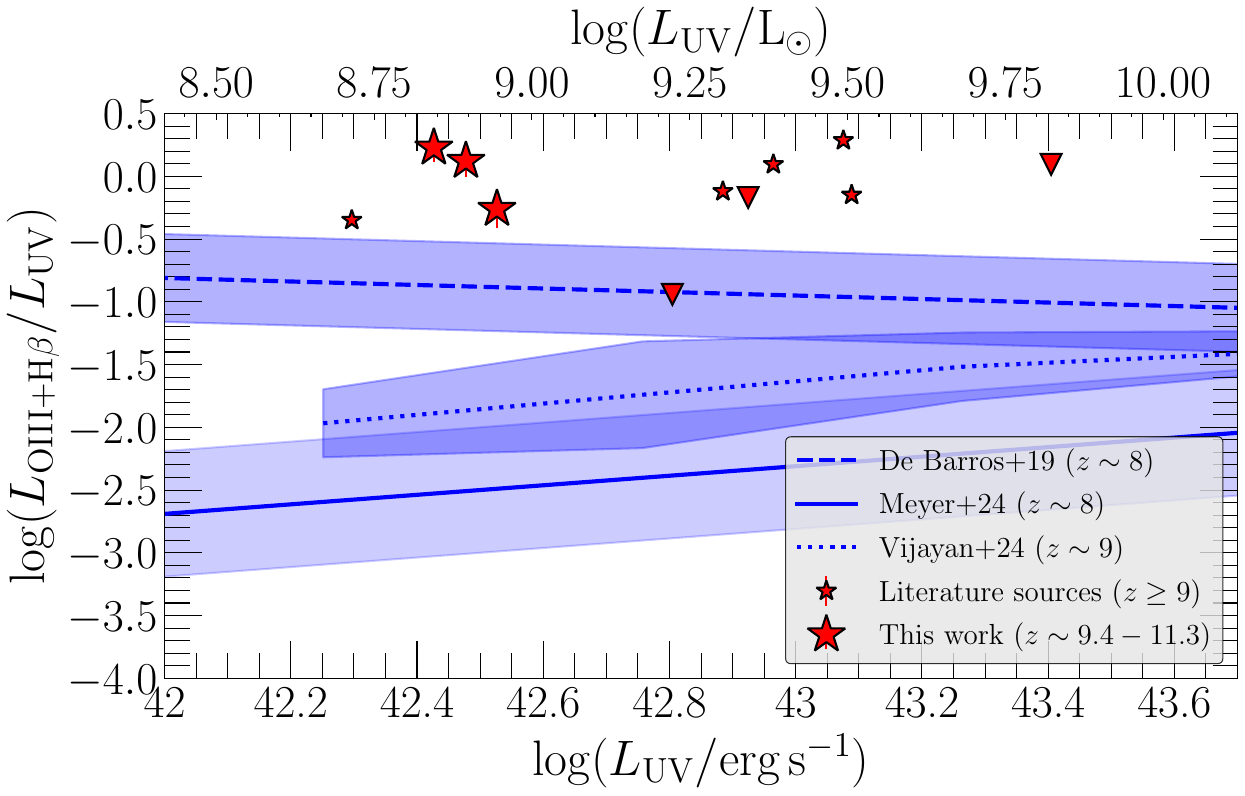}
    \caption{
        The H$\beta$+[O\textsc{iii}] to UV luminosity ratio vs UV luminosity (at $1500\,\si{\angstrom}$) for our sample (large red 
        stars) and the compiled $z\geq 9$ literature sample (small red stars and
        upper limits), except for the $z\geq 9$ PRIMAL sources \citep{Heintz25}, as
        they do not have publically available line fluxes or luminosities. Also
        shown are best-fit relations from \citet{DeBarros2019} (blue dashed line),
        \citet{Meyer2024} (blue solid line), and from the {\tt FLARES} simulations
        \citep{Lovell2021,Vijayan2024} (blue dotted line). The blue shaded regions
        indicate the r.m.s.~scatter around the fitted relations. Our sample, along
        with the $z\geq 9$ literature sample, is consistent with a flat $L_{\rm
        H\beta + [O\textsc{iii}]}/L_{\rm UV} - L_{\rm UV}$ relation and show
        $L_{\rm H\beta + [O\textsc{iii}]}/L_{\rm UV}$ ratios that are $\sim
        0.5\,{\rm dex}$ higher than the values found by \citet{DeBarros2019} (blue
        line) and more than one order of magnitude higher than the ratios derived from
        the FRESCO survey \citep{Meyer2024} and the {\tt FLARES} simulations
        \citep{Vijayan2024}.
        }
    \label{fig:LOIIIHb_LUV}
\end{figure}

In Fig.\,\ref{fig:LF} we show our $z=9.4-11.3$ luminosity function constrain
along with single-Schechter function fits based on the recent direct, unbiased
spectroscopic survey determinations of the H$\beta$+[O\textsc{iii}] luminosity
functions at $z\simeq 7$ and $\simeq 8$ \citep{Meyer2025}. These are based on
simultaneous fits to the FRESCO \citep{Meyer2024} and COSMOS-3D
\citep{Meyer2025} spectroscopically derived H$\beta$+[O\textsc{iii}] luminosity
functions. The data from those two surveys cover the line luminosity range
$\log(L_{\rm H\beta+[O\textsc{iii}]}/{\rm erg\,s^{-1}}) \simeq 41.5 - 43$ at
$z\simeq 7$ and $z\simeq 8$ (solid purple and yellow lines, respectively, in
Fig.\,\ref{fig:LF}). We also compare with the GLIMPSE NIRCam survey
\citep{Korber2025}, which derives $z\simeq 7$ and $z\simeq 8$
H$\beta$+[O\textsc{iii}] luminosity functions based on SED modelling of samples
of lensed Lyman-break galaxies (dotted purple and yellow lines, respectively,
in Fig.\,\ref{fig:LF}). GLIMPSE covers $\log(L_{\rm
H\beta+[O\textsc{iii}]}/{\rm erg\,s^{-1}}) \simeq 39 - 43$. The GLIMPSE
luminosity functions, which are derived from a similar effective survey area
($\sim 4.3-4.7\,{\rm sq.~arcmin}$) as MIDIS, generally overshoots the
luminosity functions from FRESCO/COSMOS-3D. Also, GLIMPSE shows little
evolution from $z\simeq 7$ to $8$, except at the faint end, unlike
FRESCO/COSMOS-3D, which shows significantly evolution: a $\sim 0.3\,{\rm dex}$
decrease in the luminosity function going from $z\simeq 7$ to $8$
\citep{Meyer2025}.

Our $z\simeq 9-11$ H$\beta$+[O\textsc{iii}] luminosity function estimate lies
$\sim 0.4\,{\rm dex}$ below the $z\simeq 7$ and $8$ luminosity function from
GLIMPSE, and $\sim 0.5\,{\rm dex}$ above the $z\sim 8$ luminosity function from
FRESCO/COSMOS-3D. Within the significant error bars, however, our $z\sim
9.4-11.3$ estimate is consistent with both GLIMPSE and FRESCO/COSMOS-3D. In
general, a decline in the H$\beta$+[O\textsc{iii}] luminosity is expected with
increasing redshift, given the observed evolution of the UV luminosity function
and the decreasing abundance of massive star-forming systems at $z\gs 6$
\citep[e.g.,][]{Bouwens2015, Bouwens2022, PerezGonzalez2025}. However, the
luminosity function evolution may differ from a monotonic decrease (at fixed
line luminosity) if the typical line-to-continuum ratio, $L_{\rm
H\beta+[O\textsc{iii}]}/L_{\rm UV}$, changes with redshift. At higher redshift,
harder ionising spectra and lower metallicities can increase the nebular line
output per unit UV luminosity, effectively shifting galaxies to higher $L_{\rm
H\beta+[O\textsc{iii}]}$ at fixed $L_{\rm UV}$. In this case, the
H$\beta$+[O\textsc{iii}] luminosity function may evolve more slowly than the UV
luminosity over a limited luminosity range, even if the underlying galaxy
population is rapidly declining. Even so, we would expect our  $z\simeq 9-11$
H$\beta$+[O\textsc{iii}] luminosity function estimate to lie somewhat below the
$z\sim 8$ luminosity function, as indeed is possible, given the substantial
uncertainties on our measurements due to the small number of sources and survey
area it is based on. In addition, our objects are selected via a broadband
excess consistent with strong H$\beta$+[O\textsc{iii}] emission, which
preferentially draws from the high-$L_{\rm H\beta+[O\textsc{iii}]}/L_{\rm UV}$
tail of the population. Such selection can elevate the apparent normalization
of a $1/V_{\rm max}$ LF relative to that inferred from unbiased spectroscopic
surveys, and may contribute to the differences between small-area,
line-excess-selected samples such as ours and GLIMPSE, and wide-area
spectroscopic measurements at $z\sim 7-8$ such as COSMOS-3D. Consequently,
while our data provide the first direct constraints on the normalization of the
H$\beta$+[O\textsc{iii}] LF at $z>9$, it should be viewed as indicative until
larger-area, uniformly selected spectroscopic samples become available.

Given the lack of published $z>9$ H$\beta$+[O\textsc{iii}] luminosity functions
to compare with we follow the procedure proposed by \citet{DeBarros2019} of
using robust determinations of the rest-frame $1500\,\si{\angstrom}$ UV
luminosity function and converting that to a  H$\beta$+[O\textsc{iii}]
luminosity function using an empirically derived log-linear $L_{\rm UV}-L_{\rm
H\beta+[O\textsc{iii}]}$ relation. From the $z\sim 7$ and $\sim 8$ data
presented by \citet{DeBarros2019}, we derive the following relation:
$\log(L_{\rm H\beta+[O\textsc{iii}]}{\rm / erg\,s^{-1}}) = 0.86 \log(L_{\rm
UV}{\rm /erg\,s^{-1}}) + 5.07$.\footnote{Note that \citet{DeBarros2019} derives
$\log(L_{\rm H\beta+[O\textsc{iii}]}{\rm / erg\,s^{-1}}) = 0.86 \log(L_{\rm
UV}{\rm /erg\,s^{-1}}) + 33.92$, which has a normalisation that is much higher
than the best-fit line shown in their Fig.~6.}  In comparison,
\citet{Meyer2024} derives a significantly different $L_{\rm UV}-L_{\rm
H\beta+[O\textsc{iii}]}$ relation, namely: $\log(L_{\rm
H\beta+[O\textsc{iii}]}/{\rm erg\,s^{-1}}) = 1.38 \log(L_{\rm UV}/{\rm
erg\,s^{-1}}) - 18.65$. Both relations are shown in
Fig.\,\ref{fig:LOIIIHb_LUV}, plotted as $\log(L_{\rm H\beta+[O\textsc{iii}]} /
L_{\rm UV})$ vs $\log(L_{\rm UV})$, along with the $\log(L_{\rm
H\beta+[O\textsc{iii}]} / L_{\rm UV})$ ratios of our sample (large red stars)
and the $z\geq 9$ literature sample (small red stars and upper limits). Also
shown is the relation derived from the {\tt FLARES} simulations
(\S\ref{subsection:EW-values}). We see that our MIDIS sample and the $z\geq 9$
literature sample generally exhibit very high $\log(L_{\rm
H\beta+[O\textsc{iii}]} / L_{\rm UV})$ values that are above the relation
derived by \citet{DeBarros2019} and significantly above the relations from
FRESCO \citep{Meyer2024} and the {\tt FLARES} simulations \citep{Vijayan2021}. 

We apply the three empirical $L_{\rm UV} - L_{\rm H\beta+[O\textsc{iii}]}$
relations from \citet{DeBarros2019}, \citet{Meyer2024}, and \citet{Vijayan2021}
to the $z\sim 10$ UV luminosity function from JADES \citep{Whitler2025} to
obtain predictions for the $z\sim 10$ H$\beta$+[O\textsc{iii}] luminosity
function (blue curves in Fig.\,\ref{fig:LF}). These predictions should be
interpreted as a mapping of the UV-selected population into line-luminosity
space under an assumed conditional relation $p(L_{\rm
H\beta+[O\textsc{iii}]}\,|\,L_{\rm UV})$: a super-linear (sub-linear)
conversion enhances (suppresses) the relative abundance of luminous line
emitters, thereby flattening (steepening) the predicted luminosity function.

A key point is that the appropriate conversion depends on selection. The
\citet{Meyer2024} (FRESCO) relation and the {\tt FLARES} prediction are closer
to a population-average mapping for UV-selected galaxies, whereas the
\citet{DeBarros2019} calibration is derived from samples selected via broadband
excess and therefore preferentially traces the high-$L_{\rm
H\beta+[O\textsc{iii}]}/L_{\rm UV}$ tail. In Fig.\,\ref{fig:LOIIIHb_LUV}, our
MIDIS sources (and the compiled $z\geq 9$ literature objects) occupy this
high-ratio tail, with $L_{\rm H\beta+[O\textsc{iii}]}/L_{\rm UV}$ elevated
relative to FRESCO/{\tt FLARES}. When propagated to a prediction for the
line-luminosity function, this higher line-to-continuum normalization naturally
yields a higher predicted $\Phi(L_{\rm H\beta+[O\textsc{iii}]})$ at fixed line
luminosity. Indeed, the luminosity function inferred by combining the $z\sim
10$ UV luminosity function with the \citet{DeBarros2019} conversion lies above
our MIDIS $z\simeq 9-11$ estimate in Fig.\,\ref{fig:LF}, while the {\tt
FLARES}-based conversion provides a closer match at $\log(L_{\rm
H\beta+[O\textsc{iii}]}/{\rm erg\,s^{-1}})\sim 42.5$. We emphasize that this
comparison is not expected to be exact: our MIDIS point is derived directly
from number counts via $1/V_{\rm max}$ (and is therefore independent of any
assumed $L_{\rm UV}$-to-$L_{\rm H\beta+[O\textsc{iii}]}$ conversion), whereas
the blue curves represent population-average forward models whose normalization
depends on both the adopted UV luminosity function and the assumed conversion
(and its intrinsic scatter).

\subsection{Implications for cosmic reionization}\label{section:xiion}
The large H$\beta$+[O\textsc{iii}] EWs of our MIDIS sample suggest hard,
efficient ionising spectra, making them potentially interesting candidates for
the production of the Lyman-continuum (LyC) photons required for cosmic
reionisation. A key quantity in this respect is the galaxies ionising photon
production efficiency, $\xi_{\rm ion}$. In this section, we derive the
$\xi_{\rm ion}$-values for our sample and investigate the evolution of
$\xi_{\rm ion}$ with redshift and its dependence on galaxy properties.  

\subsubsection{The ionising photon production efficiency, $\xi_{\rm ion}$}\label{subsubsection:xi_ion}
We estimate $\xi_{\rm ion}$ from the derived H$\beta$+[O\textsc{iii}] line
luminosities. Following a similar approach to that adopted by several recent
studies \citep[e.g.,][]{Matthee2023, Rinaldi2024, Heintz25}, we use the
definition $\xi_{\rm ion} = \dot{N_{\rm ion}}/L^{\rm intr}_{\rm UV, \nu}$,
where $\dot{N_{\rm ion}}$ is the intrinsic H-ionising photon production rate,
and $L^{\rm intr}_{\rm UV, \nu}$ is the intrinsic monochromatic UV luminosity
at $1500\,\si{\angstrom}$. $\dot{N}_{\rm ion}$ is related to the intrinsic
${\rm H}\beta$ luminosity through $\dot{N}_{\rm ion}(1-f_{\rm esc,LyC}) =
L^{\rm intr}_{\rm H \beta}/c_{\rm H\beta}$, where $f_{\rm esc, LyC}$ is the
escape fraction of LyC photons out of the galaxy. We assume Case~B
recombination conditions, i.e., $f_{\rm esc, LyC} = 0$, and typical conditions
($T_{\rm e} = 10^4\,{\rm K}$, $n_{\rm e} = 10^2\,{\rm cm^{-3}}$), in which case
$c_{\rm H\beta} = 4.76\times10^{-13}$ \citep[e.g.,][]{Schaerer2003}. We have
$L^{\rm intr}_{\rm H\beta} = L_{\rm H\beta} 10^{0.4 A_{\rm H\beta}}$, where
$L_{\rm H\beta}$ is the observed H$\beta$ line luminosity and $A_{\rm H\beta}$
the dust-attenuation of the line. For $A_{\rm H\beta}$, we adopt the value
obtained from our SED fits. Our F560W photometry captures the blended
H$\beta$+[O\textsc{iii}] complex, and so we convert to ${\rm H}\beta$ using the
[O\textsc{iii}]$\lambda5007$-to-H$\beta$ luminosity ratio
$r(M_{\mathrm{UV}})=L_{\rm [O\textsc{iii}]\lambda5007}/L_{\rm H\beta} =
(-1.8M_{\rm UV} - 25.7)/1.34$ for $M_{\rm UV} \in [-19, -16.5]$ and
$r(M_{\mathrm{UV}})=L_{\rm [O\textsc{iii}]\lambda5007}/L_{\rm H\beta} =
(-0.5M_{\rm UV} - 4.25)/1.34$ for $M_{\rm UV} \in [-16.5, -12.5]$
\citep[see][]{Korber2025}. Note, this calibration is based on observed line
luminosities, and so we subsequently apply the dust-correction $A_{\rm H\beta}$
to $L_{\rm H\beta}$ ($ = L_{\rm H\beta + [O\textsc{iii}]}/[1+(1 + 1/2.98)\times
r(M_{\rm UV})]$).

The intrinsic $1500\,\si{\angstrom}$ luminosity is given by $L^{\rm intr}_{\rm
UV, \nu} = L_{\rm UV, \nu}/f_{\rm esc, UV}$ where $f_{\rm esc, UV}$ is fraction
of photons escaping the galaxy in the UV continuum. We correct the UV continuum
using the \citet{Meurer1999} prescription for the \citet{Calzetti2000}
reddening law:
\begin{equation}
f_{\mathrm{esc,UV}}(\beta) =
\begin{cases}
10^{-0.83(2.23+\beta)}, & \beta > -2.23,\\
1, & \beta \le -2.23,
\end{cases}
\end{equation}
see also \citet{Rinaldi2024}. We note that while our MIDIS sample has very blue
UV continua ($\beta \ls -1.8$), none of them have $\beta < -2.23$ and thus we
cannot assume $f_{\rm{esc,UV}}\approx1$.

Uncertainties in $\xi_{\rm{ion}}$ are derived via Monte Carlo sampling of the
posterior distributions of $L_{\rm H\beta + [O\textsc{iii}]}$, $M_{\rm{UV}}$,
$\beta$, and $A_{\rm H\beta}$, propagated through the $r(M_{\rm{UV}})$ and
$f_{\rm{esc,UV}}(\beta)$ relations, and are quoted as the 16th and 84th
percentiles of the resulting distributions (see Table
\ref{tab:candidate-list}). In summary, we derive the intrinsic $\xi_{\rm ion}$,
i.e., corrected for dust attenuation for both the nebular line and UV continuum
emission. We further assume that $f_{\rm esc,LyC} = 0$ (in which case $\xi_{\rm
ion}$ is often denoted as $\xi_{\rm ion, 0}$). To test whether this is a
reasonable assumption, we derive $f_{\rm esc,LyC}$ using the prescription
provided by \citet{Chisholm2022}, who found a strong anti-correlation between
$f_{\rm esc, LyC}$ and $\beta$: $f_{\rm esc, LyC} = (1.3\pm 0.6)\times
10^{-4}\times 10^{(-1.2\pm 0.1)\beta}$, based on {\it HST} observations of
local ($z\ls 0.3$) star-forming galaxies. We find $f_{\rm esc, LyC} \ls 4\%$
for our MIDIS sources, thus justifying our assumption above. For the $z\geq 9$
literature sample $f_{\rm esc, LyC}$ varies between 0.01 and 0.38, although
12/14 sources have $f_{\rm esc, LyC} \ls 0.10$ (Table \ref{tab:xi-and-fesc}).
\begin{table}
\caption[\protect]{Ionising production efficiencies ($\xi_{\rm ion}$) for our
    MIDIS sources, as derived in \S\ref{section:xiion}, along with $xi_{\rm
    ion}$ values obtained from the literature for our $z\geq 9$ sample, see
    \S\ref{section:literature-sample} and Table \ref{tab:lit_z9_ew}. Also
    listed, are Lyman Continuum escape fractions ($f_{\rm LyC}$) as derived
    using the prescription from \citet{Chisholm2022}.
}
\centering
\renewcommand{\arraystretch}{1.2}
\begin{tabular}{l c c}
\hline
\hline
ID & $\log(\xi_{\rm ion}/{\rm Hz\,erg^{-1}})$ & $f_{\rm esc, LyC}$\\
 & [${\rm dex}$] & \\
\hline
659 &  $25.4^{+0.2}_{-0.2}$ & $0.02^{+0.01}_{-0.01}$\\
3233 & $25.4^{+0.2}_{-0.2}$ & $0.04^{+0.03}_{-0.02}$\\
3759 &  $25.1^{+0.2}_{-0.2}$ & $0.03^{+0.02}_{-0.01}$\\
\hline
PRIMAL--1       &  25.81$\pm$0.12 & $0.11^{+0.11}_{-0.06}$\\
PRIMAL--2       &  25.02$\pm$0.20 & $\cdots$\\
PRIMAL--3       &  24.70$\pm$0.18 & $0.03^{+0.02}_{-0.02}$\\
PRIMAL--4       &  25.89$\pm$0.03 & $0.09^{+0.07}_{-0.04}$\\
PRIMAL--5       &  25.47$\pm$0.02 & $0.38^{+0.44}_{-0.22}$\\
PRIMAL--6       &  25.45$\pm$0.02 &  $0.21^{+0.22}_{-0.12}$\\
PRIMAL--7       &  26.14$\pm$0.05 &  $0.03^{+0.02}_{-0.01}$\\
PRIMAL--8       &  25.50$\pm$0.03 &  $0.08^{+0.07}_{-0.04}$\\
RXJ2129--11027  &  25.60$\pm$0.11 &  $0.06^{+0.03}_{-0.03}$\\
UNCOVER--26185  &  25.50$\pm$0.06 & $0.07^{+0.06}_{-0.04}$\\
MACS0647--JD    &  25.30$\pm$0.10 & $\cdots$ \\
UNCOVER--37126  &  25.75$\pm$0.09 & $0.38^{+0.38}_{-0.21}$\\
GNz11           &  25.66$\pm$0.06 & $0.10^{+0.09}_{-0.05}$\\
CEERS2--588     &  24.90          &  $0.01^{+0.01}_{-0.01}$\\
GLASS--Z12      &  25.72          &  $0.10^{+0.08}_{-0.05}$\\
JADES-GS-z14-0  &  25.35          &  $0.03^{+0.02}_{-0.02}$\\
\hline
\end{tabular}
\label{tab:xi-and-fesc}
\end{table}

\subsubsection{The redshift evolution of $\xi_{\rm ion}$}\label{subsection:xi_z}
\begin{figure}
    \centering
    \includegraphics[width=1.0\linewidth]{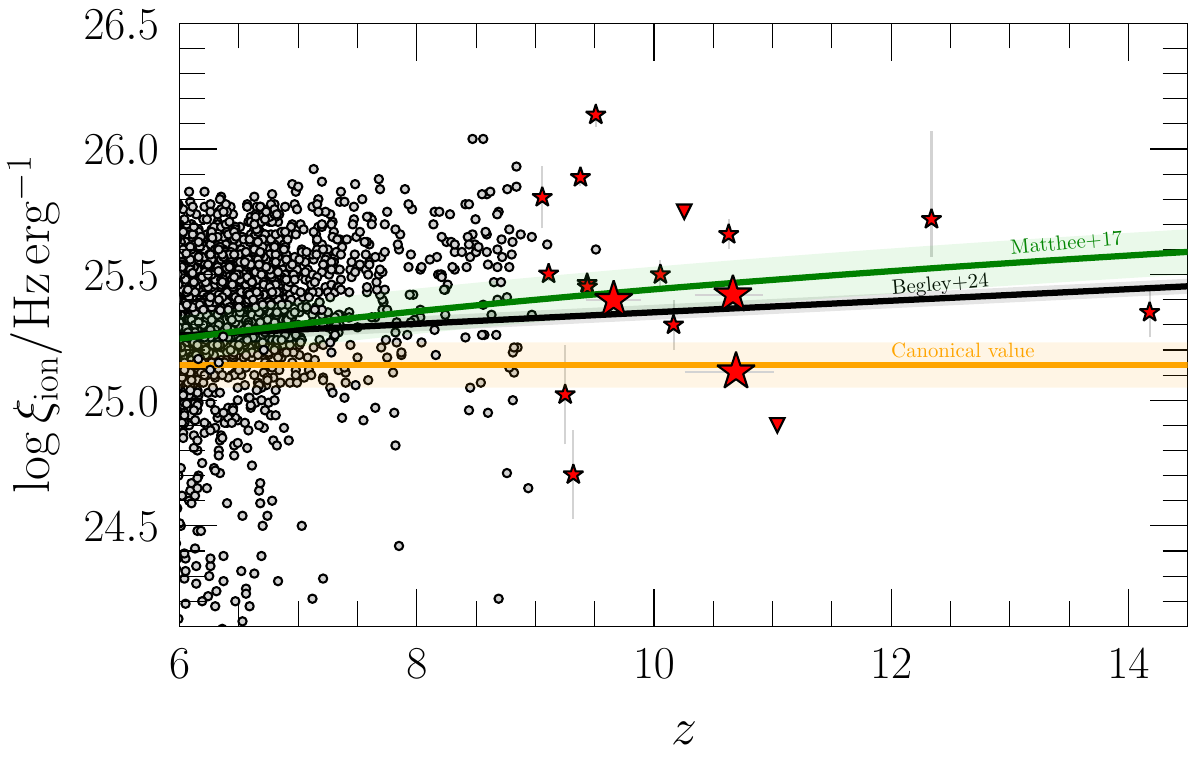}
    \caption{
        The ionising photon production efficiency, $\xi_{\rm ion}$, as a
        function of redshift for the MIDIS sources presented in this paper
        (large red stars) and the $z\geq 9$ literature sample (small red stars;
        Table \ref{tab:xi-and-fesc}). Also shown are literature samples in the
        range $z\simeq 6-9$ \citep{Stark2015, Stark2017, Stefanon2022,
        Rinaldi2024, Lin2024, Whitler2024, Simmonds2024a, Boyett2024,
        Heintz25}. The "canonical" $\log \xi_{\rm ion} = 25.2$ value for $z\gs 6$
        galaxies \citep{Robertson2013, Bouwens2016} is shown as the horisontal
        orange line, along with power-law fits to galaxy samples at $z\sim 2$
        \citep{Matthee2017} and $z\sim 7-8$ \citep{Begley2025}, shown as green and
        black curves, respectively. The broad spread at $z\geq 9$ likely reflects
        bursty SFHs and duty-cycle effects, spanning post-burst phases with weak
        recombination lines to extreme emission-line systems with very high EWs.
        }
    \label{fig:xi_vs_z}
\end{figure}

The inferred $\xi_{\rm ion}$ values for our MIDIS galaxies are shown as a
function of redshift in Fig.\,\ref{fig:xi_vs_z}, alongside the $z\ge 9$
literature compilation and comparison samples at $z\simeq 6-9$
\citep{Stark2015, Stark2017, Stefanon2022, Rinaldi2024, Lin2024, Whitler2024,
Simmonds2024a, Boyett2024, Heintz25}. Several recent studies report a positive
but relatively shallow increase of $\xi_{\rm ion}$ with redshift once sample
selection and luminosity/mass mixing are considered
\citep[e.g.,][]{Rinaldi2024, Simmonds2024a, Begley2025}. In particular,
\citet{Simmonds2024a} performed a forward-modelling null test and argue that
the observed $\xi_{\rm ion}(z)$ slope cannot be reproduced by selection effects
alone (noting that their test ignores measurement uncertainties), while
\citet{Begley2025} infer a mild redshift dependence when fitting for redshift
alone and a stronger dependence when jointly fitting for $z$ and $M_{\rm UV}$,
highlighting the importance of accounting for correlated sample properties.

We find a median $\log(\xi_{\rm ion}/{\rm Hz\,erg^{-1}})=25.40\pm0.14$ for the
MIDIS sample, placing these galaxies in the high-$\xi_{\rm ion}$ tail of the
$z\geq 9$ population. This value is consistent with the median of the compiled
$z\geq 9$ literature sample, $\log(\xi_{\rm ion}/{\rm
Hz\,erg^{-1}})=25.47\pm0.47$, which in turn is broadly consistent with typical
values reported at $z\simeq 7-8$ \citep[e.g., $\log(\xi_{\rm ion}/{\rm
Hz\,erg^{-1}})=25.55^{+0.11}_{-0.13}$;][]{Rinaldi2024}. 
MIDIS is not designed to be 

At $z\geq 9$, individual galaxies already show a wide range in inferred
$\xi_{\rm ion}$, consistent with highly time-variable star formation and a
broad range of nebular conditions. Some luminous systems 
rest-optical have values close to the ``canonical'' $\log(\xi_{\rm ion}/{\rm
Hz\,erg^{-1}})\simeq 25.2-25.3$ \citep[e.g.,][]{Hsiao2024b, Zavala2025},
whereas others show elevated efficiencies associated with very large
Balmer-line equivalent widths and extremely young stellar populations
\citep[e.g.,][]{Alvarez-Marquez26}. Conversely, the massive $z>10$ galaxy
CEERS2-588 exhibits unusually weak H$\alpha$ emission and correspondingly low
$\xi_{\rm ion}$, interpreted as a post-burst (``mini-quenching'') phase
\citep{Harikane2026}. At the extreme end, some sources have been argued to host
both high $\xi_{\rm ion}$ and substantial LyC escape fractions, implying that
rare systems could contribute disproportionately to the ionising budget
\citep[e.g.,][]{Marques-Chaves26}. Taken together, these results reinforce the
picture that the scatter in $\xi_{\rm ion}$ at $z>9$ reflects bursty SFHs and
duty-cycle effects, motivating our examination below of how $\xi_{\rm ion}$
scales with ${\rm EW}_{\rm rest}^{\rm H\beta+[O\textsc{iii}]}$, $M_{\rm UV}$
and $\beta$ in a uniform framework.

\subsubsection{$\xi_{\rm ion}$ scaling relations and dependence on galaxy properties}
\begin{figure}
    \centering
    \includegraphics[width=1.0\linewidth]{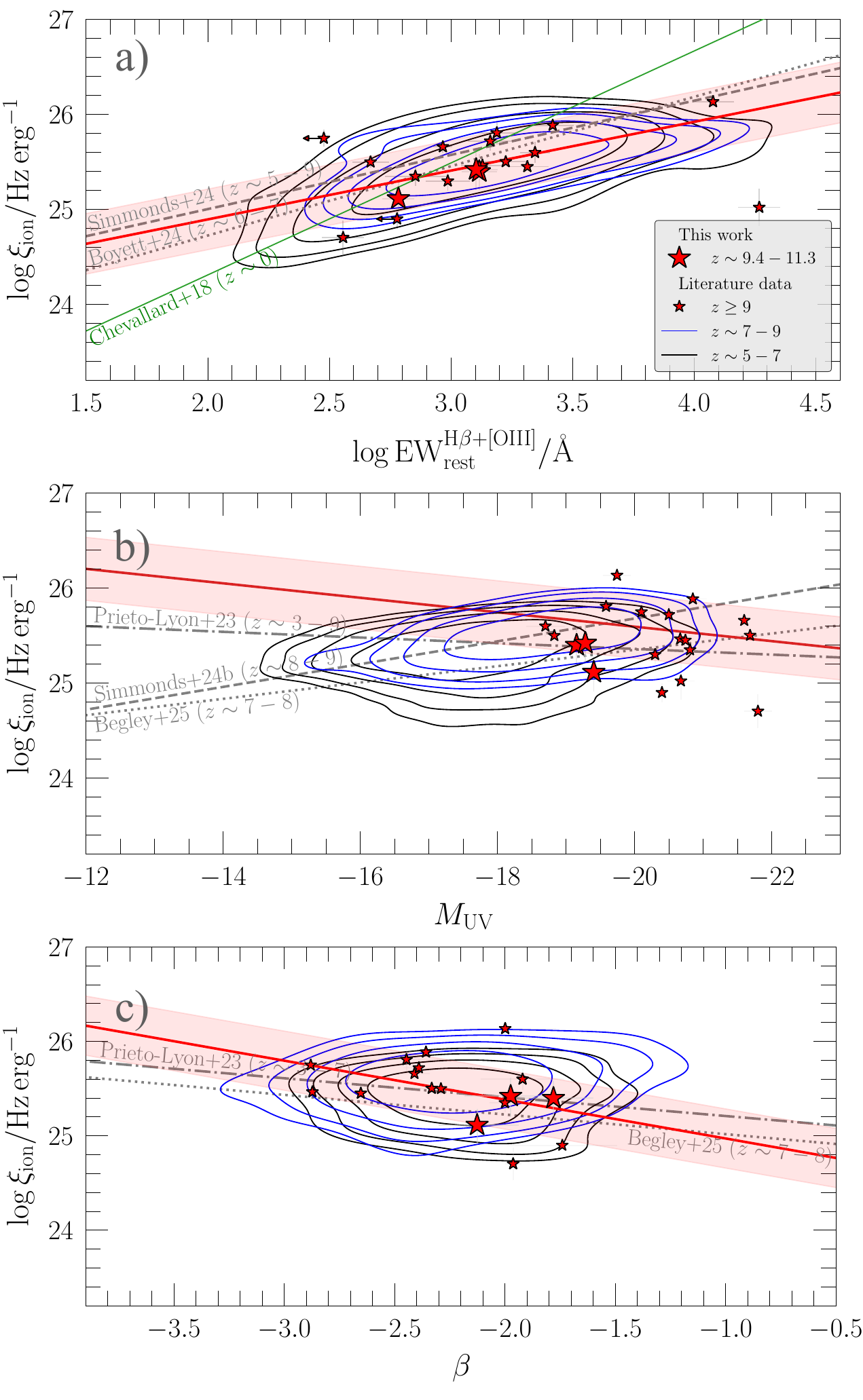}
    \caption{
        Scaling relations of $\log(\xi_{\rm ion}/{\rm Hz\,erg^{-1}})$ vs
        $\log({\rm EW}_{\rm rest}^{\rm H\beta+[O\textsc{iii}]}/\si{\angstrom})$
        (panel {\bf a)}), $M_{\rm UV}$ (panel {\bf b)}), and $\beta$ (panel
        {\bf c)}). Large red stars indicate the $z=9.4-11.3$ MIDIS sample,
        while small red stars show the $z\geq 9$ literature sample (Table
        \ref{tab:lit_z9_ew}. Literature samples at $z=5-7$ and $z=7-9$ are
        shown as black and blue contours, respectively
        \citep{Stefanon2022,Prieto-Lyon2023,Ning2023,Simmonds2023,Matthee2023,Mascia2024,Saxena2024,Lin2024,Whitler2024,
        Boyett2024,Simmonds2024b,Rinaldi2024,Heintz25,Begley2025}. In all three
        panels, the thick red line show the log-linear fit to the scaling
        relations of the $z\geq 9$ galaxies, with the residual scatter shown as
        the red shaded region. In panel a), the gray  dashed and dotted lines
        indicate translated literature relations based on spectroscopic
        measurements of ${\rm EW}^{\rm [O\textsc{iii}]\lambda5007}$
        \citep{Simmonds2024b,Boyett2024}, converted to ${\rm EW}_{\rm
        rest}^{\rm H\beta+[O\textsc{iii}]}$ using a luminosity-dependent
        prescription for the H$\beta$/[O\textsc{iii}] ratio. In panel b), the
        dashed, dotted, and dot-dashed lines show log-linear fits to large
        spectroscopic $z\gs 7$ samples from \citet{Simmonds2024b},
        \citet{Begley2025} and \citep{Prieto-Lyon2023}, respectively. 
        in panel c),  
        }
\label{fig:scalingrelations}
\end{figure}

{\it Nebular-line strength as a predictor of $\xi_{\rm ion}$}\\
Fig.\,\ref{fig:scalingrelations}a shows $\xi_{\rm ion}$ as a function of ${\rm
EW}_{\rm rest}^{\rm H\beta+[O\textsc{iii}]}$ for our $z = 9.3-11.4$ MIDIS
sample (large red stars), and the $z\geq 9$ literature sample (small red
stars). Also shown are comparison samples at $z = 7-9$ and $z = 5-7$ from the
literature \citep{Matthee2023, Boyett2024, Heintz25}. All three redshift bins
exhibit a significant $\xi_{\rm ion} - {\rm EW}_{\rm rest}^{\rm
H\beta+[O\textsc{iii}]}$ correlation (a Spearman rank correlation test yields
$p < 0.05$). 

We characterise the relation using a log-linear parameterisation of the form
$\log(\xi_{\rm ion}/{\rm Hz\,erg^{-1}}) = m\times \log({\rm EW}_{\rm rest}^{\rm
H\beta+[O\textsc{iii}]}/{\rm \si{\angstrom}}) + b$, and for the three redshift
bins, we  find the following best-fit slopes and intercepts: $(m,b)_{z\simeq
5-7} = (0.74 \pm 0.06,\, 23.13 \pm 0.21)$, $(m,b)_{z\simeq 7-9} = (0.53 \pm
0.09,\, 23.86 \pm 0.31)$, and $(m,b)_{z\geq 9} = (0.62 \pm 0.15,\, 23.60 \pm
0.47)$. Thus, within the fitting uncertainties, the derived slopes and
intercepts for the $z=5-7$, $z=7-9$ and $z\geq 9$ samples are consistent. In
Fig.\,\ref{fig:scalingrelations}a we only show the fitted relation and its
r.m.s~scatter for the $z\geq 9$ sample.

Previous spectroscopic studies of extreme [O\textsc{iii}] emitters at $z \gs 1$
have reported tight correlations between $\xi_{\rm ion}$ and ${\rm EW}_{\rm
rest}^{\rm [O\textsc{iii}]\lambda5007}$ \citep{Reddy2018,Tang2019, Boyett2024,
Simmonds2024a, Pahli2025}. Here, we limit the comparison to the studies by
\citet{Boyett2024} and \citet{Simmonds2024a}, which cover spectroscopic samples
at $z\simeq 6-7$ and $z\simeq 5-9$, respectively. To this end,  we convert
${\rm EW}_{\rm rest}^{\rm [O\textsc{iii}]\lambda5007}$ to ${\rm EW}_{\rm
rest}^{\rm H\beta+[O\textsc{iii}]}$. We do this by assuming the continuum is
approximately constant across $4861-5007\,\si{\angstrom}$, which implies that
EW ratios trace line-flux ratios, and we therefore have: ${\rm EW_{\rm
rest}^{\rm H\beta + [O\textsc{iii}]}}\simeq {\rm EW_{\rm rest}^{\rm
[O\textsc{iii}]\lambda5007}}\left(1+\frac{F_{\rm
[O\textsc{iii}]\lambda4959}}{F_{\rm
[O\textsc{iii}]\lambda5007}}+\frac{F_{\mathrm{H}\beta}}{F_{\rm
[O\textsc{iii}]\lambda5007}}\right)$, with the [O\textsc{iii}] doublet ratio
$F_{\rm [O\textsc{iii}]\lambda4959}/F_{\rm [O\textsc{iii}]\lambda5007}=1/2.98$.
As in \S\ref{subsection:xi_z}, we parametrize the term
$F_{\mathrm{H}\beta}/F_{\rm [O\textsc{iii}]\lambda5007}$ via the
luminosity-dependent prescription $r(M_{\rm UV})\equiv L_{\rm
[O\textsc{iii}]\lambda5007}/L_{\mathrm{H}\beta}$ \citep{Korber2025}, so that
$F_{\mathrm{H}\beta}/F_{\rm [O\textsc{iii}]\lambda5007}=1/r(M_{\rm UV})$. To
propagate the resulting uncertainties, we (uniformly) Monte Carlo sample
$M_{\rm UV}$ over the luminosity ranges spanned by the samples, compute the
corresponding shift $\Delta\log({\rm EW}/{\rm
\si{\angstrom}})=\log\!\left[1+1/2.98+1/r(M_{\rm UV})\right]$, and rewrite the
relation $\log(\xi_{\rm ion}/{\rm Hz\,erg^{-1}})=m\times\log({\rm
EW_{[O\textsc{iii}]\lambda5007}}/{\rm \si{\angstrom}})+b$ in terms of our EW
definition as $\log(\xi_{\rm ion}/{\rm Hz\,erg^{-1}})=m\times[\log({\rm EW_{\rm
rest}^{\rm H\beta+[\mathrm{O\textsc{iii}}]}}/{\rm
\si{\angstrom}})-\Delta\log({\rm EW}/{\rm \si{\angstrom}})]+b$. As expected,
this translation shifts the intercept ($b=23.40\pm 0.10 \longrightarrow
23.26\pm 0.10$ for \citet{Boyett2024}, and $b=23.97\pm 0.25 \longrightarrow
23.86\pm 0.26$ for \citet{Simmonds2024a}, respectively), while leaving the
slope unchanged. After translation, these two literature relations (dotted and
dashed lines in Fig.\,\ref{fig:scalingrelations}a) are consistent with our
measurements and derived $z\gs 9$ relation  within $\ls 1\sigma$.

Taken together, these results suggest that the coupling between nebular
excitation strength and ionising photon production efficiency inferred for
star-forming galaxies at $z\sim 5$, and deep into the epoch of reionization,
persists to at least $z \sim 11$.

A strong correlation is physically expected, since both quantities are a
function of the hardness of the ionising radiation field. The stellar
population age and star-formation history is likely the main driver of the
coupling, where larger EWs reflect a higher relative contribution from very
young ($\ls 10\,{\rm Myr}$), ionising stars compared to older, non-ionising
populations that dominate the continuum. As the stellar population ages,
$\xi_{\rm ion}$ drops significantly as the ionising output decreases faster
than the non-ionising UV continuum \citep[e.g.,][]{PrietoJimenez2025,
Katz2025}. The IMF is another potential important factor, where galaxies with a
top-heavy IMF harbour a higher fraction of very hot massive stars that produce
more ionising photons per UV luminosity, thus driving up $\xi_{\rm ion}$ and
${\rm EW_{\rm rest}^{\rm H\beta+[O\textsc{iii}]}}$. Dust extinction may also
play a role, especially if it msuppresses UV luminosity more than the optical
nebular lines, which would lead to an over-prediction of $\xi_{\rm ion}$
\citep{Matthee2017,Simmonds2023}. Non-negligible UV continuum escape fraction
($f_{\rm esc, UV} >> 0$) would lower both $\xi_{\rm ion}$ and the nebular line
EWs.

\smallskip

\noindent{\it UV luminosity dependence of $\xi_{\rm ion}$}\\
In Fig.\,\ref{fig:scalingrelations}b we plot $\xi_{\rm ion}$ vs $M_{\rm UV}$
for our $z = 9.4-11.3$ MIDIS sample and $z\geq 9$ literature sample. Also
shown, are comparison samples at $z = 7-9$ and $z = 5-7$ from the literature
\citep{Boyett2024, Simmonds2024b, Heintz25}.

As has been noted by recent studies, there is no clear picture regarding trends
in $\xi_{\rm ion}$ with $M_{\rm UV}$ \citep[e.g.,][]{Begley2025}. While most
studies find a weak trend with $M_{\rm UV}$, there is some disagreement in the
literature about whether the trend is for the more UV-luminous galaxies to have
slightly higher $\xi_{\rm ion}$ values on average \citep[e.g.,][]{Pahli2025,
Simmonds2024b, Begley2025} or vice versa, i.e., less UV-luminous galaxies
exhibiting higher efficiencies \citep[e.g.,][]{Prieto-Lyon2023, Simmonds2024a},
or indeed, whether there is any correlation at all
\citep[e.g.,][]{Shivaei2018}. 

For the $z=5-7$ and $z=7-9$ samples we find a modest but highly significant
anti-correlation between $\log(\xi_{\rm ion}/{\rm Hz\,erg^{-1}})$ and $M_{\rm
UV}$ (a Spearman rank correlation test yields $p < 10^{-7}$ for both samples).
A log-linear fit of the form $\log(\xi_{\rm ion}/{\rm Hz\,erg^{-1}}) = m\times
M_{\rm UV} + b$ yields $(m,b)_{z=5-7} = (-0.057\pm 0.001, 24.48\pm 0.02)$ and
$(m,b)_{z=7-9} = (-0.044\pm 0.016, 24.62\pm 0.29)$. The two relations are fully
consistent with each other, and furthermore consistent with the $z\sim 7-8$ and
$z\sim 8-9$ comparison samples from \citet{Begley2025} and
\citet{Simmonds2024b}. Thus, across the redshift range $z=5-9$ we confirm the
existence of a mild, but significant, $\xi_{\rm ion} - M_{\rm UV}$ relation in
which more UV-luminous galaxies have slightly higher $\xi_{\rm ion}$ values. We
find no statistically significant evidence for evolution in this relation over
the redshift range $z\simeq 5-9$.

For our combined MIDIS and $z\geq 9$ sample, rank correlation tests yield
results consistent with no $\xi_{\rm ion}-M_{\rm UV}$ correlation. Nominally, a
log-linear fit to the sample yields $(m,b)_{z\geq9} = (0.076\pm 0.008, 27.12\pm
0.16)$, which aligns with a trend of lower $\xi_{\rm ion}$ in more UV-luminous
systems, as proposed by \citet{Prieto-Lyon2023}, however, this is highly
uncertain given the small $M_{\rm UV}$-range of the sample. A sample, probing a
larger dynamical range in $M_{\rm UV}$ would be required to determine the
nature of how $\xi_{\rm ion}$ depends on $M_{\rm UV}$ in $z\geq 9$ galaxies.

Depending on the direction of the observed trend, different physical mechanisms
has been proposed. In a scenario, where UV-faint galaxies have higher
efficiencies, these are explained by lower-mass system having lower gas-phase
metallicities. Low metallicity environments produce harder ionising spectra,
and thus more effifient at producing ionising photons. Also, UV-faint galaxies
typically exhibit lower dust attenuation and can, therefore, maintain a higher
ratio of ionising output to observed UV light. However, given recent systematic
studies by \citet{Simmonds2024b} and \citet{Begley2025}, higher $\xi_{\rm ion}$
in UV-bright galaxies seems to be the more likely scenario. Bursty
star-formation can explain such a trend, as UV-bright galaxies are those caught
in the midst of an intense and short-lived star-formation burst, in which young
stars  produce a disproportionately large amount of ionising radiation relative
to the UV continuum. Conversely, UV-faint galaxies are likely galaxies taking a
$> 10\,{\rm Myr}$ hiatus from star-formation. Thus, their short-lived ionising
stars have died off, but the non-ionising UV continuum (which persists for
$\sim 100\,{\rm Myr}$) remains, leading to a measured drop in ionising
efficiency. Even so, a number of factors can explain the lack of an observed
strong correlation between $\xi_{\rm ion}$ and $M_{\rm UV}$. Most prominently,
UV brightening by the substantial nebular continuum in the most extreme
star-forming galaxies can systematically drive the observed $\xi_{\rm ion}$ to
lower values than their true intrinsic values \citep[e.g.][]{Katz2025}. Another
factor that might dampen the $\xi_{\rm ion}-M_{\rm UV}$ correlation is the bias
often inherent to spectroscopic surveys, where strong emission line emitters
are easier to detect, thus overestimating the fraction of high-efficiency faint
galaxies and missing the broader, non-detected population \citep{Begley2025}.
Finallly, including ${\rm EW}$, $\beta$, and $M_{\rm UV}$ in a multi-variate
analysis of $\xi_{\rm ion}$, it has been found that the $M_{\rm UV}$-dependency
largely vanishes, suggesting that $M_{\rm UV}$ is only a secondary proxy for
$\xi_{\rm ion}$, with ${\rm EW}$ and $\beta$ being the more fundamental drivers
of $\xi_{\rm ion}$ \citep{Begley2026}.

\smallskip

\noindent{\it UV continuum slope and $\xi_{\rm ion}$}\\
In Fig.\,\ref{fig:scalingrelations}c we plot $\xi_{\rm ion}$ vs $\beta$ for our
$z = 9.3-11.4$ MIDIS sample (large red stars) and the $z\geq 9$ literature
sample (small red stars). Also shown are $z = 7-9$ (blue contours) and $z =
5-7$ (black contours) comparison samples from the literature \citep{Boyett2024,
Simmonds2024b, Heintz25}. For the $z=5-7$ and $7-9$ samples we find marginally
significant $\xi_{\rm ion} - \beta$ correlations (Spearman rank coefficients of
$r_{\rm S} \simeq -0.16$, $p\simeq 0.05$). Likewise, the combined $z\geq 9$
sample shows a marginally significant anti-correlation (Spearman rank
coefficients of $r_{\rm S} \simeq -0.50$, $p\simeq 0.03$). This is consistent
with studies to date, which report a weak but significant anti-correlation
between $\xi_{\rm ion}$ and $\beta$ (e.g., \citet{Prieto-Lyon2023,
Rinaldi2024}; cf.\ \citet{Pahli2025}). These works further suggest that the
relation is largely redshift invariant, with measurements from high-redshift
photometric samples broadly consistent with those derived from larger
spectroscopic datasets \citep{Begley2025,Begley2026}. The implications of this
anti-correlation is that galaxies with bluer UV slopes tend to show somewhat
enhanced ionising photon production efficiencies. This behaviour is
qualitatively consistent with expectations since blue UV slopes ($\beta \ls
-2$) are characteristic of young, low-metallicity stellar populations with hard
ionising spectra and high LyC photon production efficiencies. Blue UV slopes
also indicate low levels of dust attenuation, which further facilitates
substantial $\xi_{\rm ion}$-values. The relatively mild anti-correlation
\citep[slope $\sim -0.2$;][]{Begley2026} likely reflects the fact that $\beta$
is influenced by multiple competing factors, including stellar population age,
metallicity, dust attenuation, and nebular continuum emission
\citep{Prieto-Lyon2023,Rinaldi2024,Begley2025}. In particular, the nebular
continuum can redden the UV slope, especially in the most extreme emitters,
which would tend to flatten the observed $\xi_{\rm ion}-\beta$ relation
\citep[e.g.,][]{Katz2025}. However, as recently shown by \citet{Begley2026},
when combined with measurements of nebular line EWs and nebular continuum
attenuation, $\beta$ is one of the most accurate predictors of $\xi_{\rm ion}$.
Nevertheless, such a calibration remains unexplored at $z\geq9$, where existing
samples are still too limited to define the joint dependence of $\xi_{\rm ion}$
on $\beta$, nebular emission-line strength, and dust attenuation in a
statistically robust manner. This would require substantially larger and more
homogeneous {\it JWST} samples than presently available.

\section{Summary and Conclusions}
Using ultra-deep MIRI imaging from the MIDIS survey in the Hubble Ultra Deep
Field, we have investigated the prevalence and properties of strong
H$\beta$+[O\textsc{iii}] emitters at $z=9.4-11.3$. Our main results can be
summarized as follows:

\smallskip

\noindent$\bullet$ We identify three galaxies at $z = 9.4-11.3$ that exhibit
significant F560W flux excesses relative to their underlying continuum. From
SED modelling we find that these excesses are consistent with strong
H$\beta$+[O\textsc{iii}] emission with rest-frame equivalent widths in the
range $\sim 600-1300\,\si{\angstrom}$ (median value $\sim
1260\,\si{\angstrom}$). The sources have UV absolute magnitudes of $-19.4 \leq
M_{\rm UV} \leq -19.2$, continuum slopes $-2.1 \leq \beta \leq -1.8$, and
stellar masses $8.0 \leq \log(M_{\star}/\Msolar) \leq 8.4$, as inferred from
SED fitting to their broadband photometry. This is similar to other strong
nebular line emitters at $z\geq 9$ published in the literature. Our findings
thus confirm the existence of galaxies exhibiting strong
H$\beta$+[O\textsc{iii}] line emission less than $500\,{\rm Myr}$ after the Big
Bang.

\smallskip

\noindent$\bullet$ We augment our MIDIS sources with a sample of 16
spectroscopically confirmed galaxies at $z \ge 9$ from the literature with
measured of H$\beta$+[O\textsc{iii}] equivalent widths and associated physical
properties. Together with our MIDIS detections, this yields a combined sample
of 19 galaxies probing rest-frame optical emission beyond $z \sim 9$. The
combined sample exhibits a median $\log({\rm EW}_{\rm
H\beta+[O\textsc{iii}]}^{\rm rest}/\si{\angstrom}) = 3.12\pm 0.17$. Splitting
the $z\geq 9$ galaxies into a UV bright and a UV faint subsample, and modeling
the underlying EW distribution as a log-normal distribution, we find tentative
evidence that the more UV luminous galaxies have larger EWs, and that the rate
of increase ($d{\rm EW}/dM_{\rm UV}\sim 164\,\si{\angstrom}/{\rm mag^{-1}}$) is
in agreement with studies at $z\simeq 6-9$ \citep{Endsley2024, Begley2025,
Begley2026}. We see no trend in the dispersion of EWs between the UV bright and
faint subsamples.

\smallskip

\noindent$\bullet$ Within our combined $z\geq 9$ (MIDIS+literature) sample, we
do not find a statistically significant anti-correlation between ${\rm EW}^{\rm
H\beta+[O\textsc{iii}]}_{\rm rest}$ and $M_{\star}$. Nonetheless, rank
correlation coefficients are consistently negative for the sample and a
log-linear fit to the data suggests an anti-correlation consistent with those
observed at lower redshift \citep[e.g.,][]{Matthee2023, Rinaldi2023}. 

\smallskip

\noindent$\bullet$ The typical equivalent widths of our $z\geq 9$ sample are
consistent with the plateau in the ${\rm EW}_{\rm rest}^{\rm
H\beta+[O\textsc{iii}]} - z$ relation established at lower redshifts. We find
no statistically significant evidence for either a dramatic upturn or a
systematic decline in H$\beta$+[O\textsc{iii}] equivalent widths at $z \geq 9$.
Instead, the observed values occupy the same region of ${\rm EW}_{\rm
rest}^{\rm H\beta+[O\textsc{iii}]}$ parameter space as galaxies at $z \sim
6-9$.

\smallskip

\noindent$\bullet$ From our three MIDIS galaxies, we place the first direct
constraint on the H$\beta$+[O\textsc{iii}] luminosity function at $z\simeq
9.4-11.3$. We find a space density of order $\Phi \sim 10^{-3.4}\,{\rm
Mpc^{-3}\,dex^{-1}}$ at $\log(L_{\rm H\beta+[O\textsc{iii}]}/{\rm
erg,s^{-1}})\sim 42.5$. Given the large Poisson and cosmic-variance
uncertainties expected for a $\sim$few arcmin$^2$ field, this value is
consistent with an overall decline relative to direct $z\sim 7-8$ measurements
\citep{Meyer2025, Korber2025}. It is also compatible with population-averaged
predictions based on observed $z\sim 10$ UV luminosity functions converted to
H$\beta$+[O\textsc{iii}] luminosity functions using $L_{\rm
H\beta+[O\textsc{iii}]}/L_{\rm UV}$ vs $L_{\rm UV}$ relations motivated by
observations and simulations.

\smallskip

\noindent$\bullet$ For our MIDIS sources, we derive ionising photon
efficiencies in the range $25.1 \leq \log(\xi_{\rm ion}/{\rm Hz\,erg^{-1}})
\leq 25.4$, consistent with $z\geq 9$ extrapolations of $\xi_{\rm ion}(z)$
relations established at lower redshifts \citep{Matthee2023, Begley2025}. The
MIDIS values are also similar to $\xi_{\rm ion}$-values derived for $z\geq 9$
sources in the literature, although there is a significant scatter in the
published ionising efficiencies \citep{Heintz25, Marques-Chaves26,
Harikane2026}. 

\smallskip

\noindent$\bullet$ We have explored the scaling relations between $\xi_{\rm
ion}$ and  ${\rm EW}_{\rm rest}^{\rm H\beta+[O\textsc{iii}]}$, $M_{\rm UV}$,
and $\beta$ for our $z\geq 9$ sample, and broadly find a consistent picture
with relations established at $z\simeq 5-9$
\citep{Prieto-Lyon2023,Simmonds2024a, Boyett2024, Begley2025, Begley2026}. In
particular, the strong correlation between $\xi_{\rm ion}$ and ${\rm EW}_{\rm
rest}^{\rm H\beta+[O\textsc{iii}]}$ and between $\xi_{\rm ion}$ and $\beta$
seem to persist at $z \geq 9$, indicating that the link between nebular
excitation, UV continuum slope and ionising efficiency is established deep into
the epoch of reionization. We find no significant relation between $\xi_{\rm
ion}$ and $M_{\rm UV}$, which is also consistent with studies at lower
redshifts.

\smallskip

Taken together, our results suggest that the physical processes governing
nebular emission in galaxies at $z \geq 9$ are broadly similar to those
operating in extreme star-forming galaxies at $z \sim 5-9$. Nebular line
measurements, either through spectroscopy or broad-band photometry, of larger
samples $z\geq 9$ is essential in order to firmly establish the distribution
and scatter in the distributions of ${\rm EW}_{\rm rest}^{\rm
H\beta+[O\textsc{iii}]}$ and $\xi_{\rm ion}$, and whether subtle evolutionary
trends with physical properties such as stellar mass, UV luminosity, continuum
slope and emerge at the highest redshifts.

\section*{Acknowledgements}
The observations analysed in this work are made with the NASA/ESA/CSA James
Webb Space Telescope (DOI:
\hyperlink{https://archive.stsci.edu/doi/resolve/resolve.html?doi=10.17909/z7p0-8481}{10.17909/z7p0-8481}).
TRG, SG and IJ acknowledge funding from the Cosmic Dawn Center (DAWN), funded
by the Danish National Research Foundation (DNRF) under grant DNRF140. TRG and
IJ are also grateful for support from the Carlsberg Foundation via grant
No.~CF20-0534. Some of the data products presented herein were retrieved from
the Dawn {\it JWST} Archive (DJA). DJA is an initiative of the Cosmic Dawn
Center (DAWN), which is funded by the Danish National Research Foundation under
grant DNRF140. PGP-G acknowledges support from grant PID2022-139567NB-I00
funded by Spanish Ministerio de Ciencia, Innovaci\'on y Universidades
MICIU/AEI/10.13039/501100011033, and the European Union FEDER program {\it Una
manera de hacer Europa}. G\"O, AB and JM acknowledge support from the Swedish
National Space Administration (SNSA). LC acknowledges support from the “la
Caixa” Foundation (ID 100010434), fellowship code LCF/BQ/PR24/12050015. LC and
JAM acknowledge support from grant PID2021-127718NB-100 funded by SPanish
Minsiterio de Ciencia, Innovación y Universidades
MICIU/AEI/10.13039/501100011033, and the European Union FEDER program Una
Manera de hacer Europa. ACG acknowledges support by {\it JWST} contracts
B0215/{\it JWST}-GO-02926 and B0354/{\it JWST}-GO-08051. JAM acknowledges
support by grant PIB2021-127718NB-100 funded by MCIN/AEI/10.13039/501100011033
and by ERDF `A way of making Europe'. JPP and TVT acknowledge financial support
from the UK Science and Technology Facilities Council, and the UK Space Agency.
JH and DL were supported by research grants (VIL16599, VIL54489) from VILLUM
FONDEN. SEIB is supported by the Deutsche Forschungsgemeinschaft (DFG) under
Emmy Noether grant number BO 5771/1-1. For the purpose of open access, the
authors have applied a Creative Commons Attribution (CC BY) licence to the
Author Accepted Manuscript version arising from this submission. The data
products presented herein were retrieved from the Dawn {\it JWST} Archive
(DJA). DJA is an initiative of the Cosmic Dawn Center, which is funded by the
Danish National Research Foundation under grant No. 140. We are grateful to
Dr.~Kasper Heintz for sharing the PRIMAL catalog with us prior to it becoming
public. We are also grateful to Dr.~Charlotte Simmonds for making providing us
with the data catalogs published in \citet{Simmonds2024b}. We thank Dr.~Aswin
Vijayan for making the {\tt FLARES} simulation data-set available and for
clarifying how the line emission is implemented in the simulations. We are
grateful to Dr.~Francesco Valentino and Dr.~Minju Lee for fruitful discussions
on the paper. TRG is also grateful to Silje Liu Greve for encouragement and
inspiring discussions. This paper is dedicated to the memory of Dr.~Hans Ulrik
N\o rgaard-Nielsen and his decades-long contribution to DTU Space and Danish
astronomy.



\bibliographystyle{mnras}
\bibliography{bib} 

\bsp	
\label{lastpage}
\end{document}